\documentclass[12pt,preprint]{aastex}

\usepackage{epsfig}
\usepackage{graphicx}
\usepackage{ifthen}
\usepackage{amssymb}
\usepackage{subfigure}
\def\kms{\,\rm km\,{s}^{-1}}
\def\msun{M_\odot}
\def\Nb{{N_{\rm ngb}}}
\def\Np{{N_{\rm p}}}
\def\Mp{{M_{\rm p}}}
\def\beq{\begin{equation}}
\def\eeq{\end{equation}}
\def\pppm{\,\rm P^3M}
\def\omega0{\Omega_{\rm m,0}}
\def\kpc{\,\rm kpc}
\def\mpc{\,\rm Mpc}

\begin{document}

\title{Smoothing Algorithms and High-order Singularities in Gravitational Lensing}
\author{Guo-Liang Li\altaffilmark{1,4}, S. Mao\altaffilmark{2},
  Y.P. Jing\altaffilmark{1}, X. Kang\altaffilmark{1},
 M. Bartelmann\altaffilmark{3}}
\altaffiltext{1}{Shanghai Astronomical Observatory; the Partner Group
of MPA, Nandan Road 80, Shanghai 200030, China; Email:
{\tt (lgl,ypjing, kangx)@center.shao.ac.cn}}
\altaffiltext{2}{ University of Manchester, Jodrell Bank Observatory,
  Macclesfield, Cheshire SK11 9DL, UK; Email: {\tt smao@jb.man.ac.uk}}
\altaffiltext{3}{Zentrum f\"ur Astronomie, ITA, Universit\"at
  Heidelberg, Albert-\"Uberle-Str. 2, 69120 Heidelberg, Germany; Email:
{\tt (mbartelmann)@ita.uni-heidelberg.de}}
\altaffiltext{4}{Graduate School of Chinese
Academy of Sciences, Beijing 100039, China}
\shorttitle{}
\shortauthors{Li et al.}

\begin{abstract}
We propose a new smoothing method for obtaining surface densities from
discrete particle positions from numerical simulations. This is an essential step for many applications
in gravitational lensing. This method is based on the ``scatter''
interpretation of the discrete density field in the Smoothed Particle Hydrodynamics.
We use Monte Carlo simulations of uniform density fields and one
isothermal ellipsoid to empirically derive the noise
properties, and best smoothing parameters (such as the number of nearest neighbors
used). A cluster from high-resolution simulations is then used to
assess the reality of high-order singularities such as swallowtails and
butterflies in caustics, which are important for the interpretation of
substructures in gravitational lenses. 
We also compare our method with the Delaunay tesselation field estimator
using the galaxy studied by Bradac et al. (2004), and find good agreements.
We show that higher order singularities are not only
connected with bound subhaloes but also with the satellite streams.
However, the presence of high-order singularities are sensitive to
not only the fluctuation amplitude of the surface density, but also
the detailed form of the underlying smooth lensing potential (such as ellipticity and external shear).
\end{abstract}

\section{INTRODUCTION}

Gravitational lensing provides an important technique
to study the matter distribution in the universe (e.g., Kochanek, Schneider \&
Wambsganss 2004). On galaxy-scales, simple smooth isothermal
models appear to match the gravitational lenses reasonably
well (e.g., Kochanek 1991), although there are some difficulties
in accurately reproducing the observed flux ratios. At present, it is unclear
what causes these ``anomalous flux ratios''
(e.g., Kochanek \& Dalal 2005; Mao et al. 2004; Macci\`o et al. 2006;
Macci\`o \& Miranda 2006).
On cluster scales, it became clear early on that smooth spherical
models under-predict the number of giant arcs by orders of magnitude
(e.g., Bartelmann \& Weiss 1994; Bartelmann et al. 1995; Bartelmann et al. 1998).
Ellipticity and substructures in clusters have dramatic effects
on the lensing cross-sections of giant arcs. This is because many massive clusters are
more or less super-critical, so a small change in the surface
density or the shear can have a large nonlinear effect on the lensing properties.

The effects of ellipticity and substructures are difficult to
model analytically. The most realistic way to account for these is to use
numerical simulations. $N$-body simulations can accurately track the
evolution of dark matter particles to highly non-linear
regimes. State-of-the-art simulations can resolve dark matter haloes
with several million (or more) particles within the virial radius. Even at such
high resolutions, each particle in the simulation is only a Monte Carlo
realization of the underlying (smooth)
density field.  To predict the lensing properties of the discrete
particle distributions, the first step is to obtain an accurate smoothed surface
density field. The fluctuations of the smoothed surface density maps
directly affect the shapes,
magnifications and orientations of images, and as a consequence, the
lensing cross-sections and optical depths. In addition, high-resolution $N$-body
simulations indicate that there are also many substructures in galaxies and
clusters of galaxies. A good smoothing algorithm must be able to
obtain a surface density map with sufficiently low
noise level while retaining the information of substructures.

Previous workers have proposed many different algorithms for smoothing.
Bartelmann et al. (1998, hereafter B98) used a three-dimensional
Gaussian kernel to calculate the density on a three-dimensional grid and then integrate
along the line of sight to obtain the surface density. Bradac
al. (2002) directly obtained the surface density using
a Gaussian kernel with a fixed smoothing length
on the projected particle distributions. Li et al. (2005)
and Macci\`o et al. (2006) used adaptive kernels commonly used in the
Smoothed Particle Hydrodynamics (SPH) with similar numbers of neighbors
to obtain the surface density. The most complex method was used by
Bradac et al. (2004)
who obtained the volume density using the Delaunay
tessellation field estimator (Schaap \& van de Weygaert 2000).
and then projected the 3D density onto a 2D grid. Other
methods can also operate on a 2D particle distribution with simple smoothing
kernels, such as the nearest grid points, cloud-in-cell, and
triangular shaped clouds (see Bartelmann 2003 for a review).

It is straightforward to show that for smoothing kernels (such as
Gaussian kernels) with a fixed smoothing length, the same results are obtained
if one directly smoothes the 2D projected particle distribution or
if one first obtains the three-dimensional density distribution and then
performs the projection. Furthermore, the noise levels from these two methods
are identical. Therefore, if one uses smoothing kernels with a fixed
length, there is no need to perform the smoothing in 3D, one can simply
perform the smoothing on the projected particle distributions,
saving both memory and CPU time.
Of course, the particles in the $N$-body simulations are not
really infinitesimal points, they in fact represent Monte Carlo
realizations of a small but
finite volume in the phase space. The volume a particle occupies is
inversely proportional to the local mass density: in high density
regions, the particles occupy a smaller volume and vice versa.
Adaptive smoothing is therefore essential to
account for the large variation in the local densities.

In general, the noise level is higher if one first projects the particle
positions to 2D and then smoothes the density rather than first smoothing
the  density in 3D and then performing the projection.
This is because in the first method,
when we use adaptive smoothing (e.g., for a fixed number of
neighbors in SPH), there is an extra noise contribution due to chance
alignments. This effect is particularly serious for low-density regions.
In addition, substructures will be more difficult to identify in the
projected particle distributions. It is therefore important
to adaptively smooth the density field in 3D and perform the projection afterwards, in order
to reduce the noise level and retain the substructures. The Delaunay tessellation field estimator
used by Bradac et al. (2004) is a significant step, but the method,
as the authors pointed out, is quite time-consuming. To improve the
computational efficiency, Ascalibar \& Binney (2004) proposed a new
method, FiEstAS, which can calculate the phase-space density
distribution (in 6D) efficiently. However, in this method, there is some
ambiguity about the smoothing length and kernel.
In this paper, we concentrate on the SPH smoothing in 3D, and show that
it can be used to obtain low-noise surface density maps while at the
same time retaining the substructures. In \S2, we discuss our method. In
\S3, we study the noise properties of the method for a uniform density field and
an isothermal ellipsoid; analyses of noise properties
for these simple cases have not yet been performed in great detail
in the literature
(see Lombardi \& Schneider for a general discussion). We then take a cluster from cosmological
simulations of structure formation, and study the reality of high-order
singularities, and how they are connected with
substructures and satellite streams.  As shown by Bradac et al. (2004),
these tend to produce large flux anomalies compared with
simple analytical predictions, so it is crucial to check
what produced these and whether they are real and not from numerical
noise.  For this purpose, we also present a detailed comparison
between our method and the Delaunay tessellation method
(Schaap \& van de Weygaert 2000) used
by Bradac et al. (2004) for their simulated galaxy. In \S4, we discuss the implications of
our results on the detections of substructures in lensing galaxies and
clusters.

\section{METHOD}

The central idea of the new smoothing algorithm is to use the three-dimensional density field information,
but smooth the density in two-dimensions with SPH kernels. (Our method also
applies to Gaussian smoothing kernels, as long as we choose the
smoothing length accordingly.) Specifically, we first choose the
smoothing length according to the number of neighbors ($\Nb$) in
3D. The SPH smoothing kernel is given by
\begin{equation}
  W(r)={\sigma \over h^\nu}\left\{\begin{array}{ll}
    1-{3 \over 2}({r \over h})^2+ {3 \over 4}({r \over h})^3    & \mbox{for}\quad 0 \le {r \over h} \le 1 \\
    {1 \over 4}{(2-{r \over h})}^3 & \mbox{for}\quad 1 < {r \over h} \le 2 \\
    0                     & \mbox{otherwise}
	       \end{array}\right.\;,
\label{eq:19}
\end{equation}
where $\nu$ is the number of dimensions, $\sigma$ is a normalization with the values:
$2/3$, $10/7\pi$, $1/\pi$ in one, two and three dimensions respectively,
and $2h$ is the radius of a sphere (or a circle in 2D and
half-length in 1D) which contain all the neighbours.
This method is fully adaptive:
for high-density regions, the smoothing length will be smaller and vice
versa. Once the smoothing length is determined, we distribute the
particle mass onto the surrounding grid points. The weights are derived from the SPH kernel in two dimensions, using the particle's smoothing length and its distance from the grid point in the projection plane. This corresponds to the
``scatter'' method in calculating the density in SPH, see Hernquist \&
Katz (1989). The smoothed surface density
on a grid point along any projection can then be obtained by simply adding up the
contributions of all the grid points that have non-vanishing
contributions. We refer our method as the ``scatter and integrate'' (SI)
method. This method uses three-dimensional density information, but as the
partition of the particle weight onto the grid points is performed
only once, it can be implemented efficiently.

\section{RESULTS}

In the following, we will apply our method and examine its performance in three cases.
1) a uniform density field, in order to understand the noise property of
this method;
2) an isothermal ellipsoid; in this case,
we will examine how different smoothing methods reproduce the well-known analytical
critical curves and caustics. We will also check whether noise can produce
high-order singularities such as swallowtails. 3) A simulated
galaxy cluster taken from Jing \& Suto (2002). For this cluster,
we will check how well the (sub)structures are preserved in the smoothed
surface density maps, and how they affect high-order
singularities and the cusp relation in different smoothing methods in \S \ref{sec:cluster}.
 In \S3.4, we make detailed comparisons of our method with the Delaunay tessellation
method for the simulated galaxy used by Bradac et al. (2004).

\subsection{Uniform density field}

We randomly generate $\Np$ point particles in a cube with a unit
sidelength. We use $\Np=6000, 48000, 384000$, and 3072000; they roughly
span the range found from low-resolution to high-resolution simulated clusters
within the virial radius. For each simulated cube,
we then obtain the volume density $\hat{\rho}$ at 4000 random
positions in 3D using the SPH kernel with $\Nb=32, 128, 256, 512$, and the surface
density $\hat {\kappa}$ on a $50\times 50$ grid using the SI method we discussed
above and the standard 2D SPH smoothing with the same number of
neighbors, $\Nb$. Obviously both the volume and surface densities
 have an expectation value of unity, $\bar\rho=1$ and $\bar\kappa=1$.

\begin{figure}
  \begin{center}
    \mbox{
      \subfigure[]{\scalebox{0.4}{\includegraphics{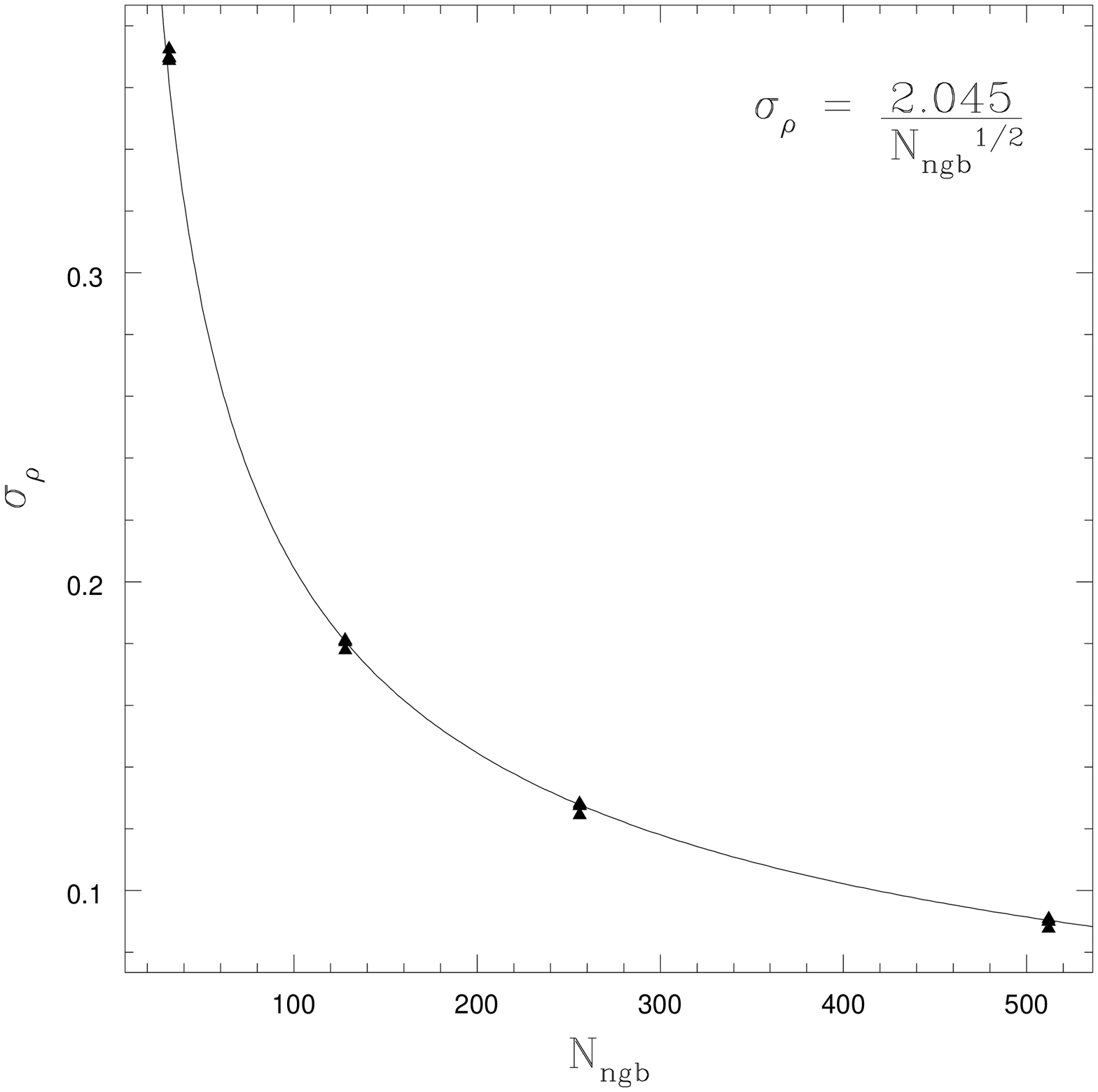}}} 
      \quad
      
      \subfigure[]{\scalebox{0.4}{\includegraphics{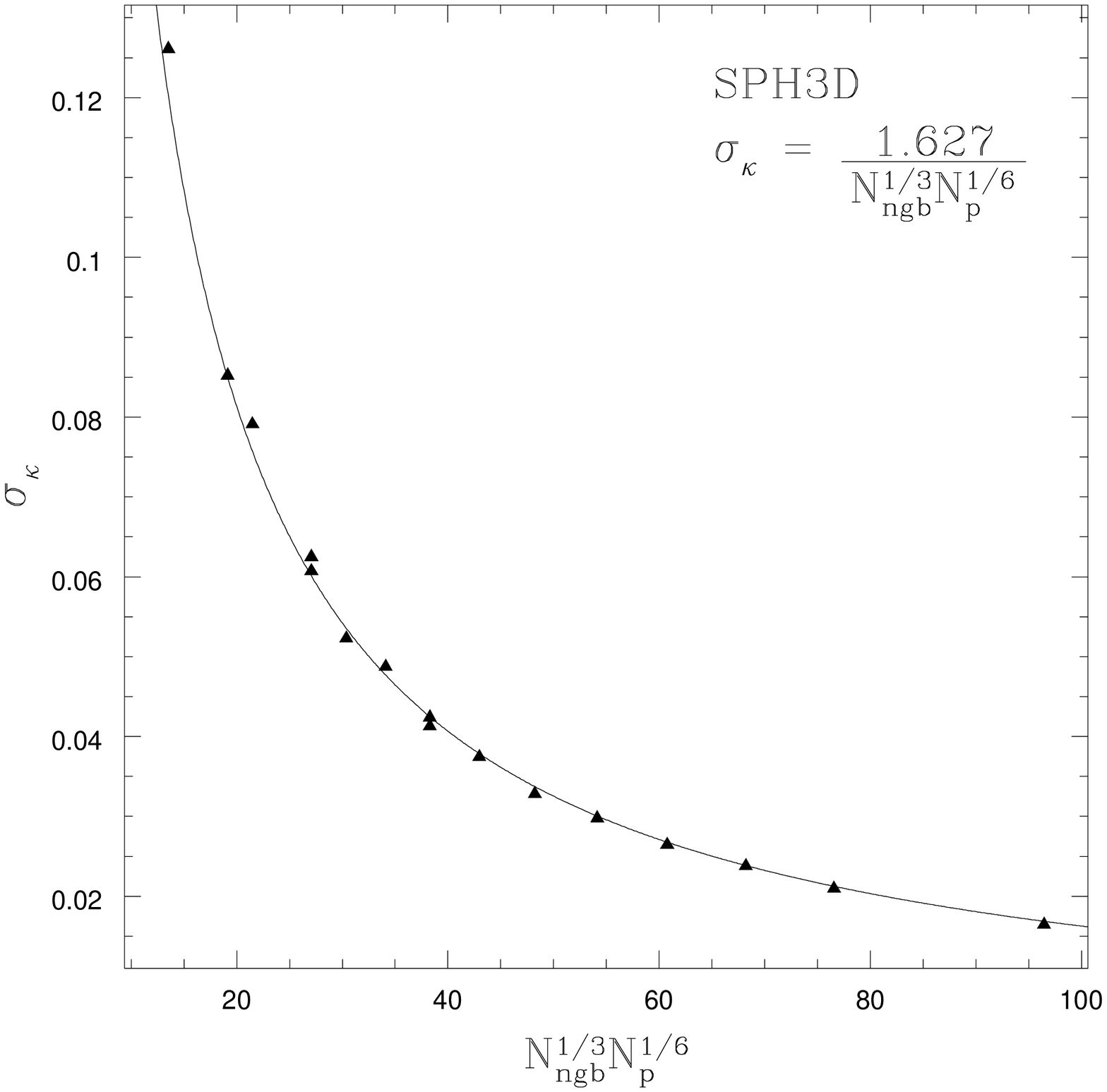}}} 
      \quad
      }
  \end{center}
  \caption{The left panel shows the noise levels of
 the volume density derived using the SPH kernel in 3D for a uniform volume density field.
 The different points correspond to different realizations.
    The solid curve is our empirically derived noise formula, eq. (\ref{eq:2d}).
The right panel shows  the noise levels for the surface density using
the scatter and integrate (SI) algorithm. 
    The solid curve is our empirically derived noise formula, eq. (\ref{eq:SI}).
  }
  \label{fig:formula}
\end{figure}

The left panel in Fig. \ref{fig:formula} shows the noise behavior for
the volume density. The standard deviation of the volume density can be well fitted as
\beq
{\sigma_\rho\over \bar\rho} \approx {2 \over \sqrt{\Nb}}.
\eeq
If we smooth the density field using SPH kernels directly in the projected particle
distribution, we find that
\beq
{\sigma_\kappa \over \bar\kappa}={1.698 \over \sqrt{\Nb}}
\approx
 \sqrt{{2 \over 3}} ~~
{\sigma_\rho \over \bar{\rho}}
\label{eq:2d}
\eeq
The factor $\sqrt{2/3}$ arises because the noise contribution along the
line of sight has been eliminated due to integration.

If we use the scatter and integrate (SI) method discussed above, we find that
the noise scales as
\beq
\left({\sigma_\kappa \over \bar\kappa}\right)_{\rm SI}={1.627 \over \Nb^{1/3} \Np^{1/6}}
\label{eq:SI}
\eeq
This fitting formula matches the noise properties in our simulations
quite well, as can be seen in Fig. \ref{fig:formula}.

It may appear that eqs. (\ref{eq:2d}) and (\ref{eq:SI}) are very
different. In fact, for the same fixed smoothing length, these two
formulae are related. In the following arguments, to avoid confusion, we label the number of neighbor in 2D and 3D explicitly. For an $N_{\rm ngb3d}$ neighbor in 3D, the smoothing length
scales as $r_{\rm s} \propto {N_{\rm ngb3d}}^{1/3} \Np^{-1/3}$. The number of neighbors in 2D will
be $N_{\rm ngb2d} \propto r_{\rm s}^2 \Np \propto {N_{\rm ngb3d}}^{2/3}\Np^{1/3}$. Putting this into
eq. (\ref{eq:2d}) yields $\sigma_\kappa/\bar{\kappa} \propto {N_{\rm ngb3d}}^{-1/3}
\Np^{-1/6}$, identical to the dependence on $\Np$ and $\Nb$ as given in eq. (\ref{eq:SI}).
For a non-uniform field, however, our method will yield a
more exact local density since the chance alignments due to projections are eliminated and,
more importantly, the information of substructures and voids is better kept.
We will illustrate these in \S\ref{sec:nbody}.

\label{sec:uniform}

\subsection{Isothermal ellipsoids \label{sec:SIE}}

Isothermal ellipsoids (e.g., Kormann, Schneider \& Bartelmann 1994;
Keeton \& Kochanek 1998) are reasonable models for galaxy-scale lenses.
They have analytical critical curves and caustics, so they offer
an ideal model to test and calibrate different smoothing algorithms.

We generated many realizations of an oblate isothermal cluster. Its
physical density is given by (cf. eq. 2 in Keeton \& Kochanek 1998)
\begin{equation}
\rho = {{v_{\rm c}}^2 \over 4 \pi G q_3} {e \over \sin^{-1} e} {1 \over
  s^2+R^2 + z^2/q_3^2},
\end{equation}
where $v_{\rm c}$ is the circular velocity of the halo, $s$ is the core
radius, $q_3$ is the ratio of the minor axis to the other two equally long
axes, and $e=\sqrt{1-q_3^2}$. We take $v_{\rm c}$ to be
1500\,$\kms$, $s=20$\,kpc, and $q_3=0.5$. The cluster is
viewed along one of the two major axes, resulting in elliptical
isophotes with an axial ratio of 0.5. We generate particles
within a cube of sidelength of $4\mpc$. Below we present the results
from two realizations. The low-resolution simulation has a total of
$\Np=54760$ particles in the box, each with mass
 $\Mp=1.6 \times 10^{10} h^{-1}M_\odot$, while
 the high-resolution simulation has
$\Mp=0.05\times 10^{10} h^{-1}M_\odot$ and $\Np=1748982$. Here, $h$ is the Hubble constant in units of $100\kms\mpc^{-1}$.  They roughly
span low-resolution and high-resolution clusters typically found in
numerical simulations.
The redshift of the lens and source are taken to be 0.3 and 1.0
respectively, and we assume the $\Lambda$CDM cosmological model (see \S\ref{sec:cluster}).

 Different $\kappa$ maps on a 2048$\times$ 2048 grid are produced
 by the 2D SPH and SI method for
these two Monte-Carlo simulations. The lensing potential is then obtained
using the fast Fourier transform method (Bartelmann et al. 1998). To resolve the
critical curves well, we perform cubic spline interpolations of
the lensing potential on a finer grid (Li et al. 2005). At the same time, we
derive the magnification and the lens mapping from the
image plane to the source plane on the whole grid. To locate the
critical curves, we use the fact that they separate regions of
different signs of magnifications in the image plane (see, e.g., Bartelmann
2003 for a practical implementation).

\begin{figure}
  \begin{center}
    \mbox{
      \subfigure[]{\scalebox{0.4}{\includegraphics{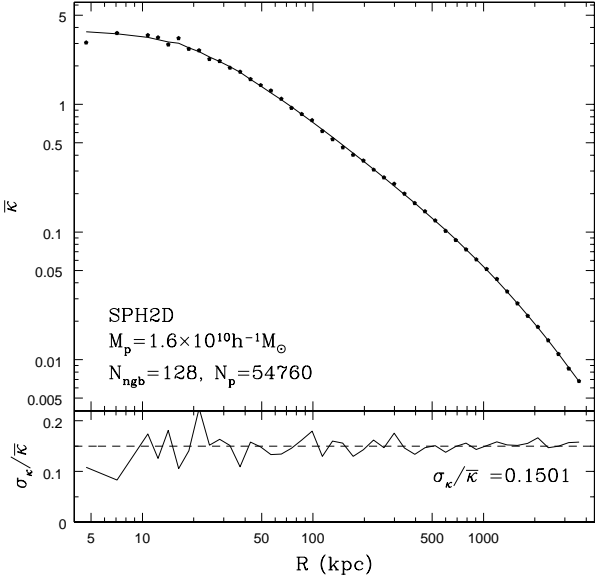}}} 
      \quad
      
      \subfigure[]{\scalebox{0.4}{\includegraphics{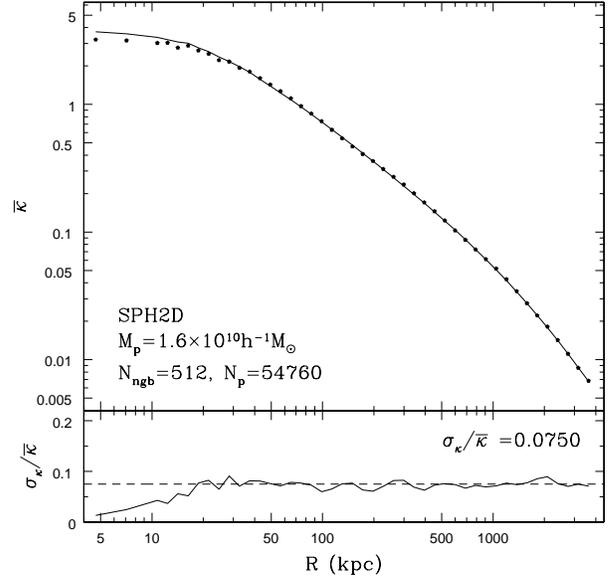}}} 
      \quad
    }
    \mbox{
      \subfigure[]{\scalebox{0.4}{\includegraphics{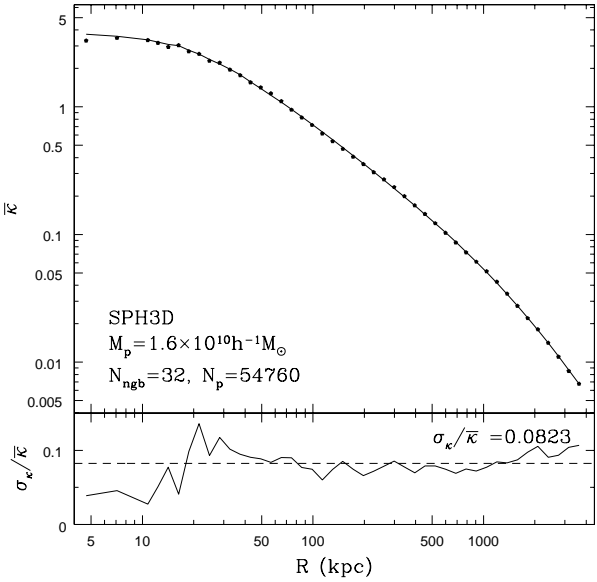}}} 
      \quad
      \subfigure[]{\scalebox{0.4}{\includegraphics{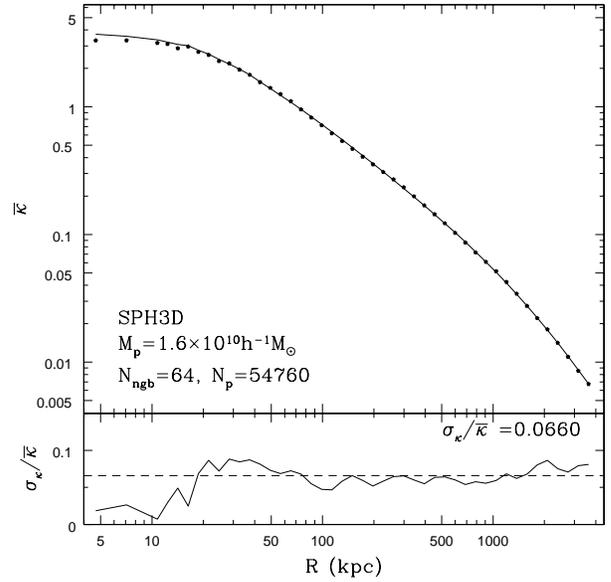}}} 
      \quad
    }
  \end{center}
  \caption{
    The surface density $\kappa$ and relative standard deviation for the
    isothermal ellipsoid model as a function of the
    ellipsoidal radius $R$. The particle mass, the smoothing method and the    
    number of particles used in smoothing are labelled in each panel. The solid curve
    shows the analytical surface density.
  }
  \label{fig:shapem1.6}
\end{figure}

Fig. \ref{fig:shapem1.6} shows the surface density obtained for the
low-resolution simulation for
different smoothing algorithms and parameters together with the
predicted analytical surface density. The surface density
is plotted as a function of the ellipsoidal radius.  As can be seen,
the analytical surface density is reproduced quite well by both the 2D SPH and the SI method,
except in the inner-most regions where the numerical results are lower due to smoothing.

The noise in the surface density is obtained from the fluctuations in the surface
density of all pixels within each elliptical annulus. For the 2D SPH smoothing, this is a function of the
number of neighbors only, and is independent of $\Np$ or the distribution
of the surface density. Eq. (\ref{eq:2d}) appears to match the noise in the
data quite well (see the top panels in Fig. 2). On the other hand, for the SI method, the surface
density is obtained by an integration of the volume density along the
line of sight. While the relative error of the volume density only depends on
the number of neighbors used, the absolute error depends on the
local density through the integration along the line of sight.
 From the lower panels in Fig. \ref{fig:shapem1.6}, we see that
the noise level changes with radius, but it appears that
eq. (\ref{eq:SI}) still provides a
reasonable approximation to the errors except in the innermost regions.
However, the noise  in $N$-body simulations is not completely Poissonian because the particles follow
dynamical evolution and their phases are correlated. Studies about the errors in $N$-body simulations
(Monaghan 1992; Niedereiter 1978) show that they are generally
smaller than a naive estimate from pure Poissonian noise. Nevertheless,
we believe our empirical formula
eq. (\ref{eq:SI}) still provides an approximate measure of the noise in the surface
density from cosmological $N$-body simulations.

\begin{figure}
  \begin{center}
    \mbox{
      \subfigure[]{\scalebox{0.4}{\includegraphics{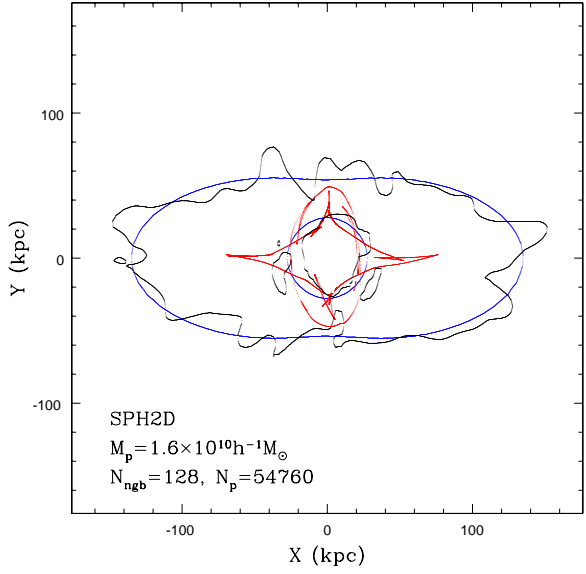}}} 
      \quad
      
      \subfigure[]{\scalebox{0.4}{\includegraphics{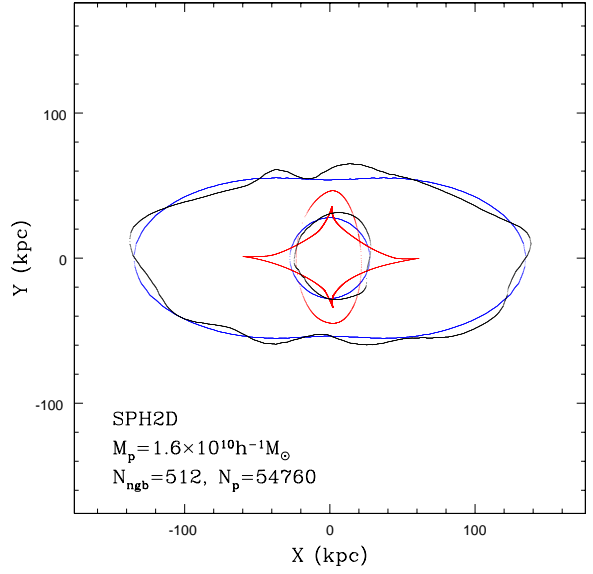}}} 
      \quad
    }
    \mbox{
      \subfigure[]{\scalebox{0.4}{\includegraphics{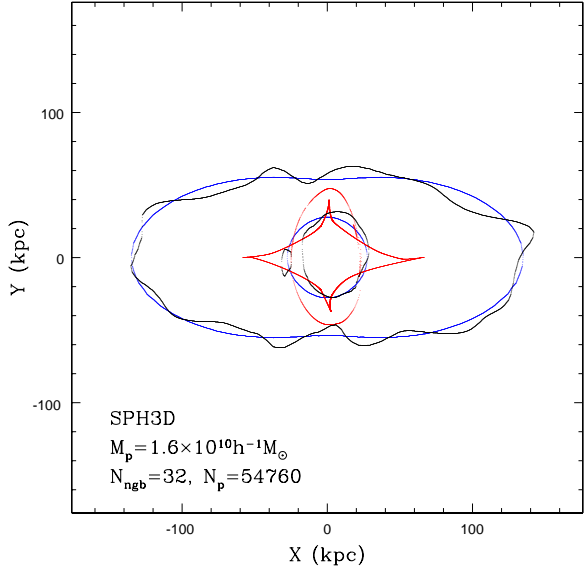}}} 
      \quad
      \subfigure[]{\scalebox{0.4}{\includegraphics{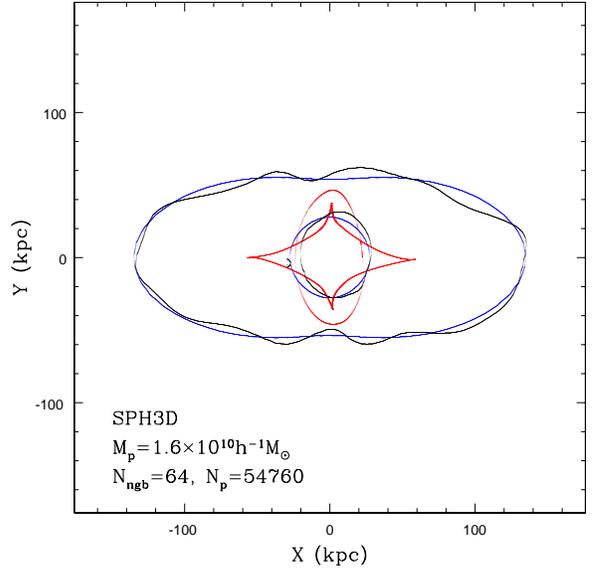}}} 
      \quad
    }
  \end{center}
  \caption{   
    The caustics and critical curves in the SI method and the 2D SPH
    smoothing method. The particle mass and the particle
 number used to smooth are indicated for each. The blue lines are the 
analytic critical curves, the black and red lines are the 
critical curves and caustics obtained from the Monte Carlo simulations.
  }
  \label{fig:critm1.6}
\end{figure}

Fig. \ref {fig:critm1.6} shows the resulting critical curves and caustics for the
smoothed density field of the low resolution realization. It is clear that for the 2D SPH smoothing, when
$\Nb=128$, the critical curves show large deviations from the analytical
results, and the caustics show a large number of high-order
singularities, such as swallowtails and butterflies.
The agreement is substantially better when $\Nb$ is increased
to 512 in the 2D SPH smoothing, in this case all the prominent higher-order
singularities disappear (in agreement with analytical predictions). This
demonstrates that these are due to artifacts of inappropriate smoothing.
For the SI method, with $\Nb=32$, the
critical curves and caustics already resemble the analytical results quite well,
with no apparent high-order singularities. Increasing $\Nb$ by a factor of 2 does not change the critical curves
and caustics substantially.

\begin{figure}
  \begin{center}
    \mbox{
      \subfigure[]{\scalebox{0.4}{\includegraphics{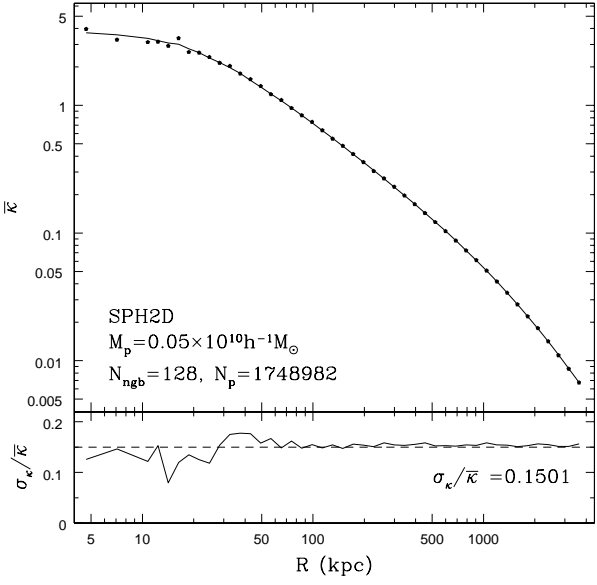}}} 
      \quad
      
      \subfigure[]{\scalebox{0.4}{\includegraphics{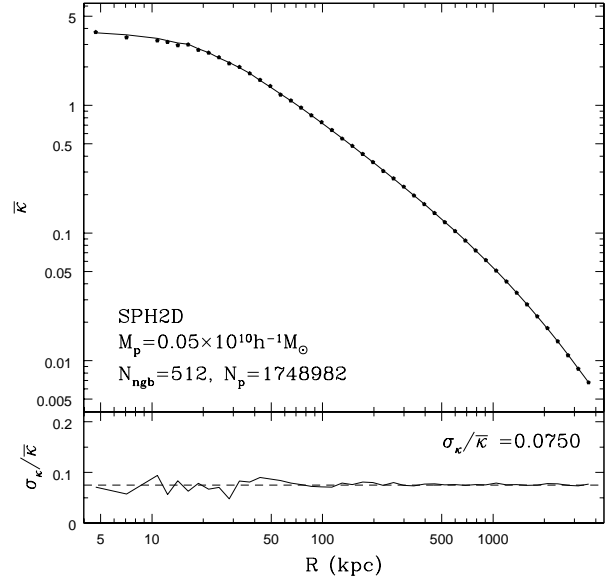}}} 
      \quad
    }
    \mbox{
      \subfigure[]{\scalebox{0.4}{\includegraphics{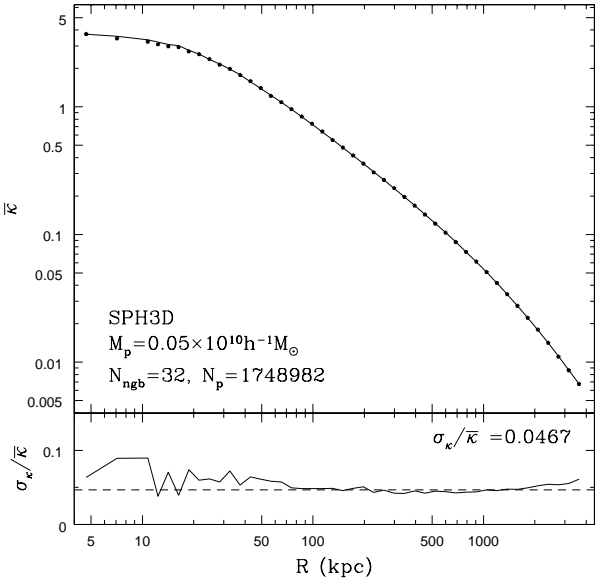}}} 
      \quad
      \subfigure[]{\scalebox{0.4}{\includegraphics{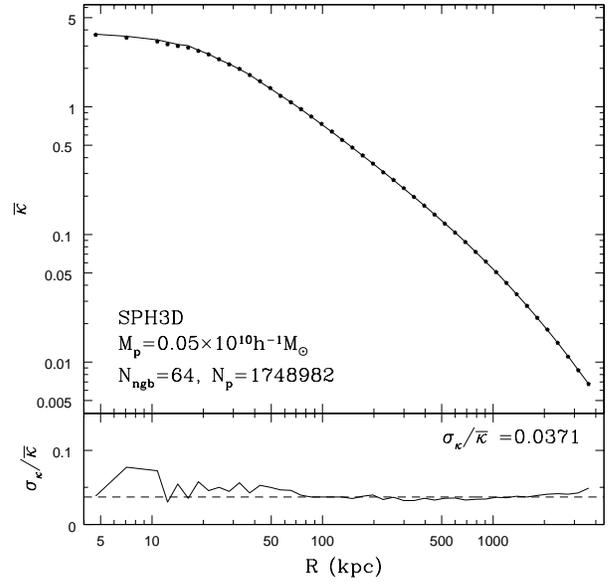}}} 
      \quad
    }
  \end{center}
  \caption{Same as Fig. 2, but for a particle mass of $0.05 \times 10^{10} h^{-1} M_\odot$ and a total
number of particles, $\Np=1748982$.
}
  \label{fig:shapem0.05}
\end{figure}

\begin{figure}
  \begin{center}
    \mbox{
      \subfigure[]{\scalebox{0.425}{\includegraphics{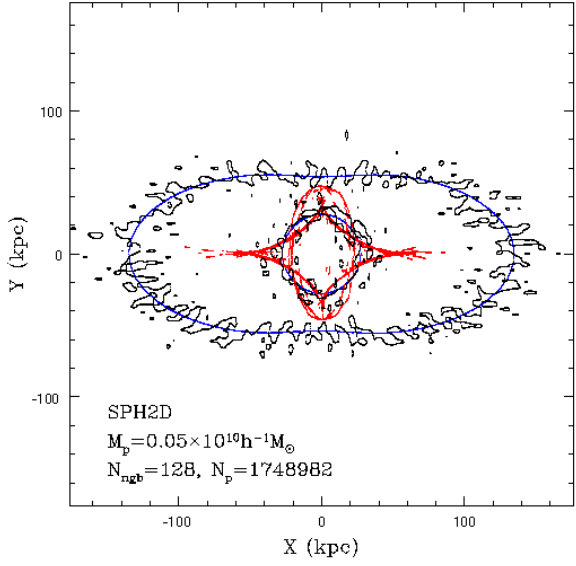}}} 
      \quad
      
      \subfigure[]{\scalebox{0.425}{\includegraphics{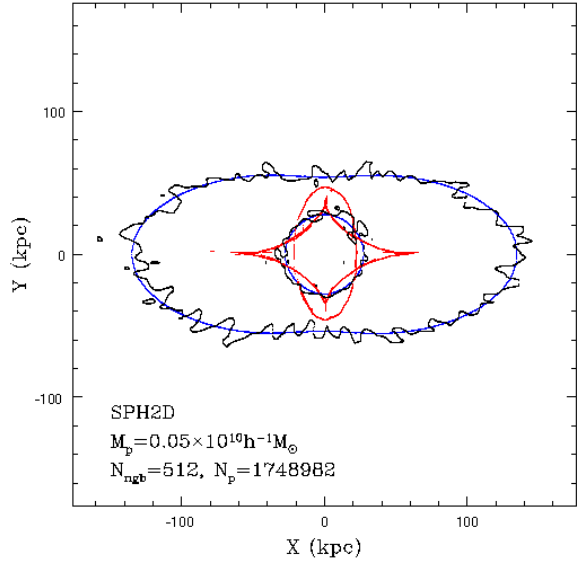}}} 
      \quad
    }
    \mbox{
      \subfigure[]{\scalebox{0.4}{\includegraphics{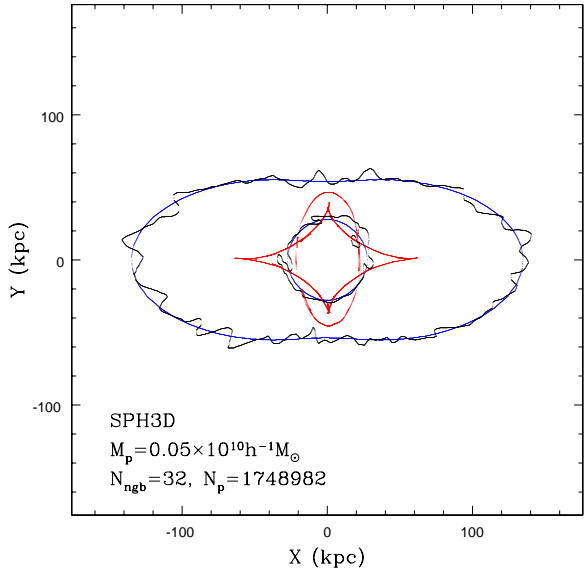}}} 
      \quad
      \subfigure[]{\scalebox{0.4}{\includegraphics{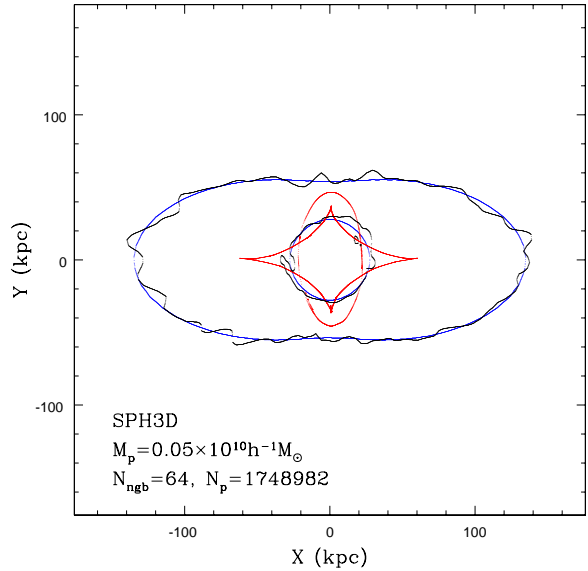}}} 
      \quad
    }
  \end{center}
  \caption{Same as Fig. 3, but for a particle mass of
 $0.05 \times 10^{10} h^{-1} M_\odot$ and a total
number of particles, $\Np=1748982$. A high resolution version is available at 
http://202.127.29.4/mppg/english/data/.
}
  \label{fig:critm0.05}
\end{figure}

The results are similar for the high-resolution Monte Carlo
simulation. Compared with the lower resolution, the surface density
(Fig. \ref{fig:shapem0.05})
is reproduced well even in the inner-most regions due to the
increased particle number. Again the 2D SPH smoothing algorithm appears to
produce artificial high-order singularities (see
Fig. \ref{fig:critm0.05}) for $\Nb=128$;
some remain even for $\Nb=512$. In contrast, the SI method does not
produce prominent higher-order singularities.

The SI method is also computationally efficient. For the
low resolution cluster, the speed for the SI method is similar to
that for the 2D SPH smoothing method. But for the
high resolution simulation, for the same noise level, the SI method
is much faster, as the 2D SPH smoothing algorithm requires many more
neighbors to be taken into account.

\subsection{A cluster from numerical simulations and its high-order
  singularities \label{sec:cluster}}

As we mentioned before,
the fluctuations in the Monte Carlo simulations discussed above for the uniform
density field and the isothermal ellipsoid arise due to Poisson
statistics. In $N$-body simulations, the noise may be significantly
smaller than the Poisson noise. Furthermore, substructures cannot be
properly accounted for in our simple simulations. To remedy this, we
study a numerically simulated cluster.

\subsubsection{Numerical simulations, substructures and streams}

The cosmological model considered here is the current popular
$\Lambda$CDM model with the matter-density parameter $\Omega_0=0.3$ and the
cosmological constant $\Lambda_0=0.7$ ($\Lambda$CDM). The shape parameter
$\Gamma=\Omega_0 h$ and the amplitude $\sigma_8$ of the linear density
power spectrum are taken to be 0.21 and 0.9, respectively, where
$h=0.7$ is again the Hubble constant in units of $100\kms\mpc^{-1}$.
A cosmological $N$-body
simulation with a box size $L=100h^{-1} \mpc$, which was generated with a
vectorized-parallel $\pppm$ code (Jing \& Suto 2002; Jing 2002), is used in
this paper. The simulation uses $512^3$ particles, so the particle
mass $m_p$ is $6.2\times 10^{8}h^{-1}\msun$. The gravitational force is
softened with the $S2$ form (Hockney \& Eastwood 1981) with the
softening parameter $\eta$ taken to be $3h^{-1}\kpc$.
Notice that this simulation has higher resolution than the one used in Li et
al. (2005), which has $\eta=30h^{-1}\kpc$ and a  mass resolution of
$1.67\times 10^{10}h^{-1} M_\odot$.

A cluster with a virial mass of $M=3.4 \times 10^{14} h^{-1} M_\odot$ and
a virial radius of $1.6h^{-1}\mpc$  at
redshift 0.326 is selected from this simulation for illustrative purposes.
And we assume the source is at redshift 2.
(this redshift is chosen so that we have large and well-resolved caustics).
We consider the projected surface mass density in a cube
with sidelength of $4\mpc$. The lens center is taken to be the center of the
cluster and there are $\Np=565600$ particles in the cube.
Bound substructures are found using the algorithm {\tt SUBFIND}
(Springel et al 2001).
This method does not find unbound substructures,
such as disrupted satellite galaxies which may manifest themselves as
satellite streams (e.g., Sagittarius in the Milky Way).

In principle, at least two different algorithms can be used to identify these streams and investigate their influence
on higher-order singularities. One is to trace the particles in the
phase space (e.g., Helmi et al. 2003; Ascasibar \& Binney 2005),
while the other is explicitly following
the evolution of accreted satellite galaxies. We tried both
methods, but the results from these two methods are similar, so we
present mainly those from the method where the
substructures are identified in the phase space (see below for details). The
second (trace-back) method revealed that the cluster consists
of two subclusters at redshift unity, which eventually merge to produce the cluster we see
at $z=0.3$. The virialization in the cluster
formation is however incomplete, as can also be seen from the density
contours in Fig. \ref{fig:contours}.

\begin{figure}
  \begin{center}
    \mbox{
      \subfigure[]{\scalebox{0.4}{\includegraphics{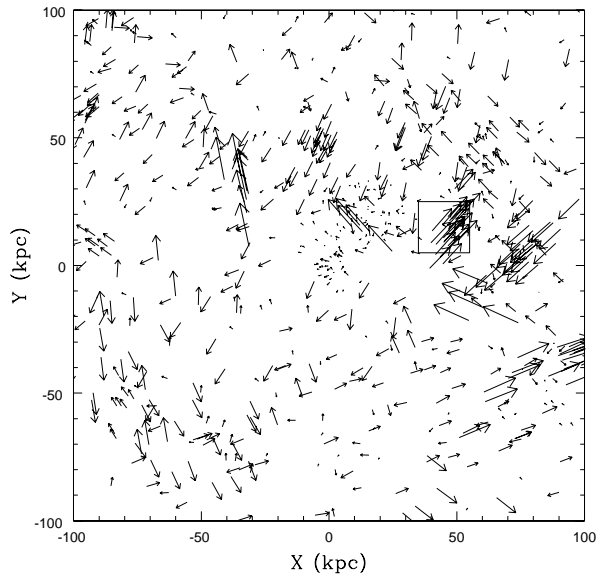}}}
    }
  \end{center}
  \caption{Streams identified in the phase space. There are 690
    particles in this region, about 1\% of the particles in the plotted
    region. The small box at (45, 15) kpc indicates 
the stream that significantly affected one of the higher-order singularities shown
    in the Fig. \ref{fig:caustics}b and c.
  }
  \label{fig:streams}
\end{figure}

To identify the satellite streams in the phase space, we first removed the bound
substructures we found from the cluster using {\tt SUBFIND}. All the substructure particles within the lensing box are
excluded. The volume density of each particle,
{$\rho(i)$} is then calculated with the SPH method using 64 neighbors. A characteristic length is defined as
${\rho(i)}^{-1/3}$. Then the Friends-Of-Friends (FOF) algorithm is used to find the
structures in the phase space. We find that the density of particles in velocity space is
approximately constant as a function of radius,
hence a constant linking length in the velocity of $0.2\sigma$
is appropriate, where $\sigma$ is the velocity dispersion for all
the particles relative to the cluster center.
In the real space, the volume density changes rapidly as a function of
position, so we use a variable linking length, $2\times {\rho(i)}^{-1/3}$.
Such a linking length allows us to extract a stream that
has a volume density a factor of $2^3$ smaller than the local volume
density, but discards streams with a much lower amplitude. Our method can easily find
coherent short-wavelength structures but cannot extract very low-amplitude
large-scale features in the phase space.
Fig. \ref{fig:streams} shows the streams we
have found: there are 690 particles in the region with a total mass
of $4.28\times 10^{11} h^{-1} M_\odot$. This is about 1\% of the
projection mass in the plotted region ($200\kpc \times 200 \kpc$) in
Fig. \ref{fig:streams}.

\subsubsection{Surface mass density maps, critical curves and caustics}

\begin{figure}
  \begin{center}
    \mbox{
      \subfigure[]{\scalebox{0.32}{\includegraphics{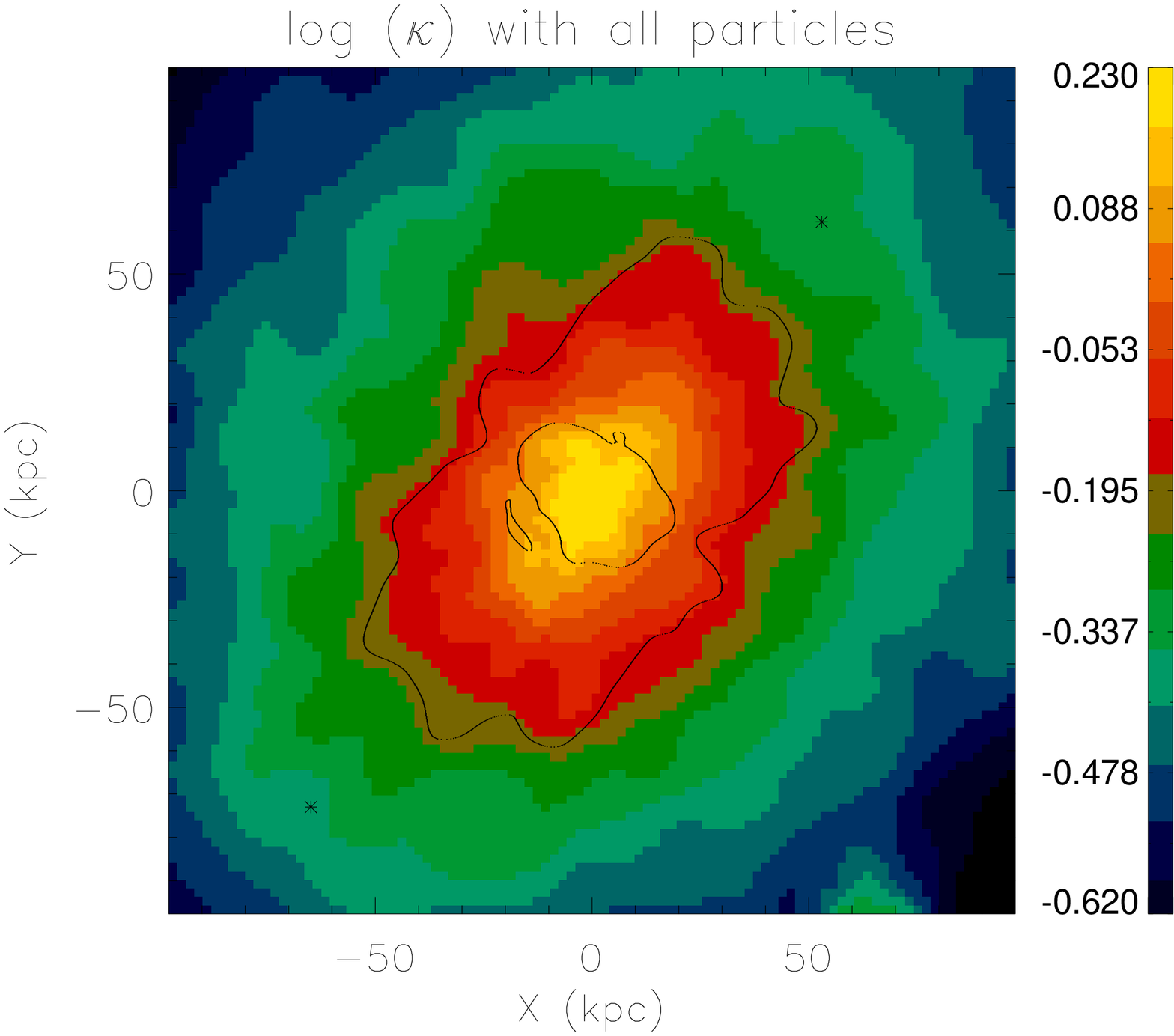}}} 
      \quad
     }
    \mbox{
      \subfigure[]{\scalebox{0.32}{\includegraphics{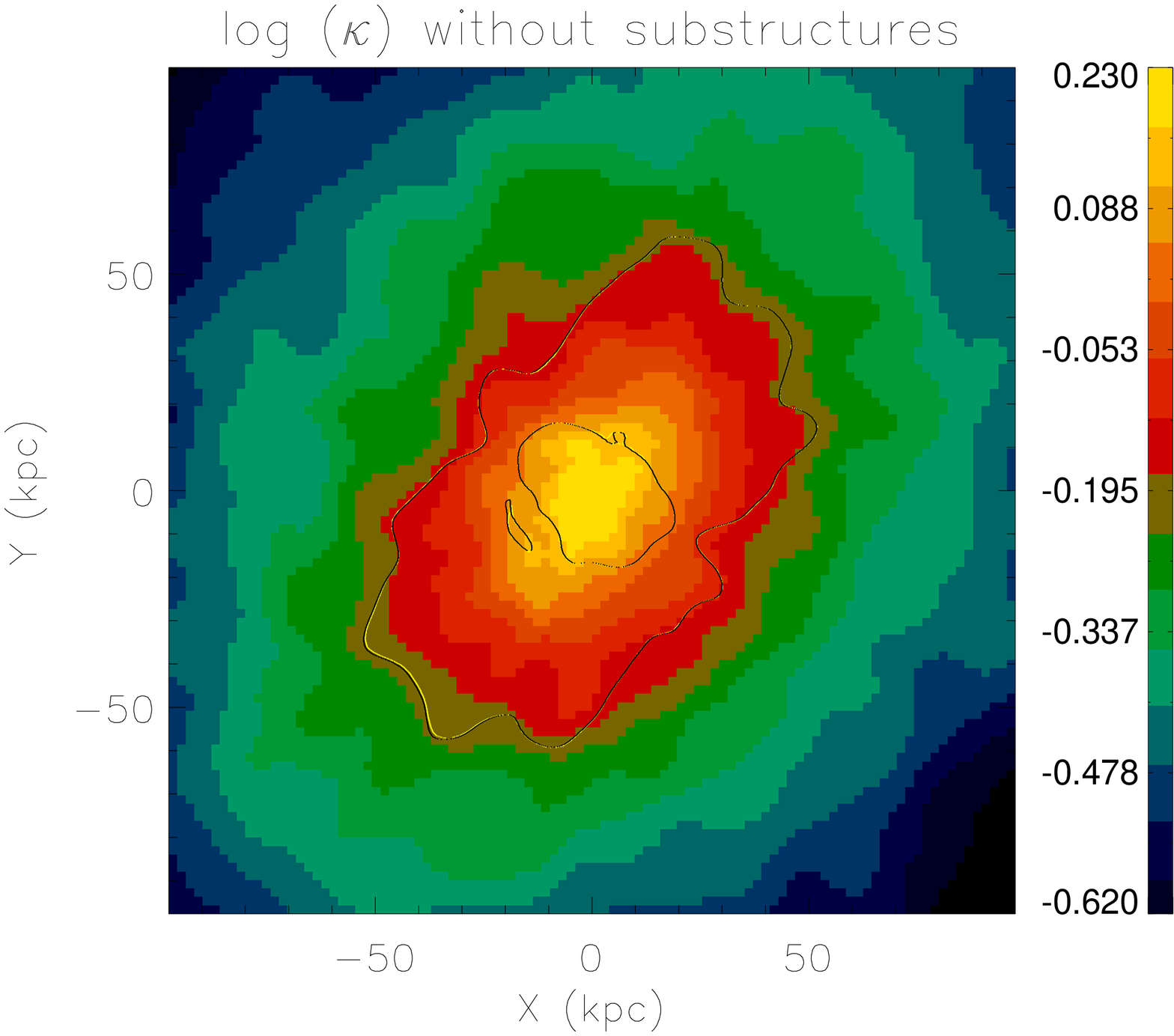}}} 
      \quad
    }
    \mbox{
      \subfigure[]{\scalebox{0.32}{\includegraphics{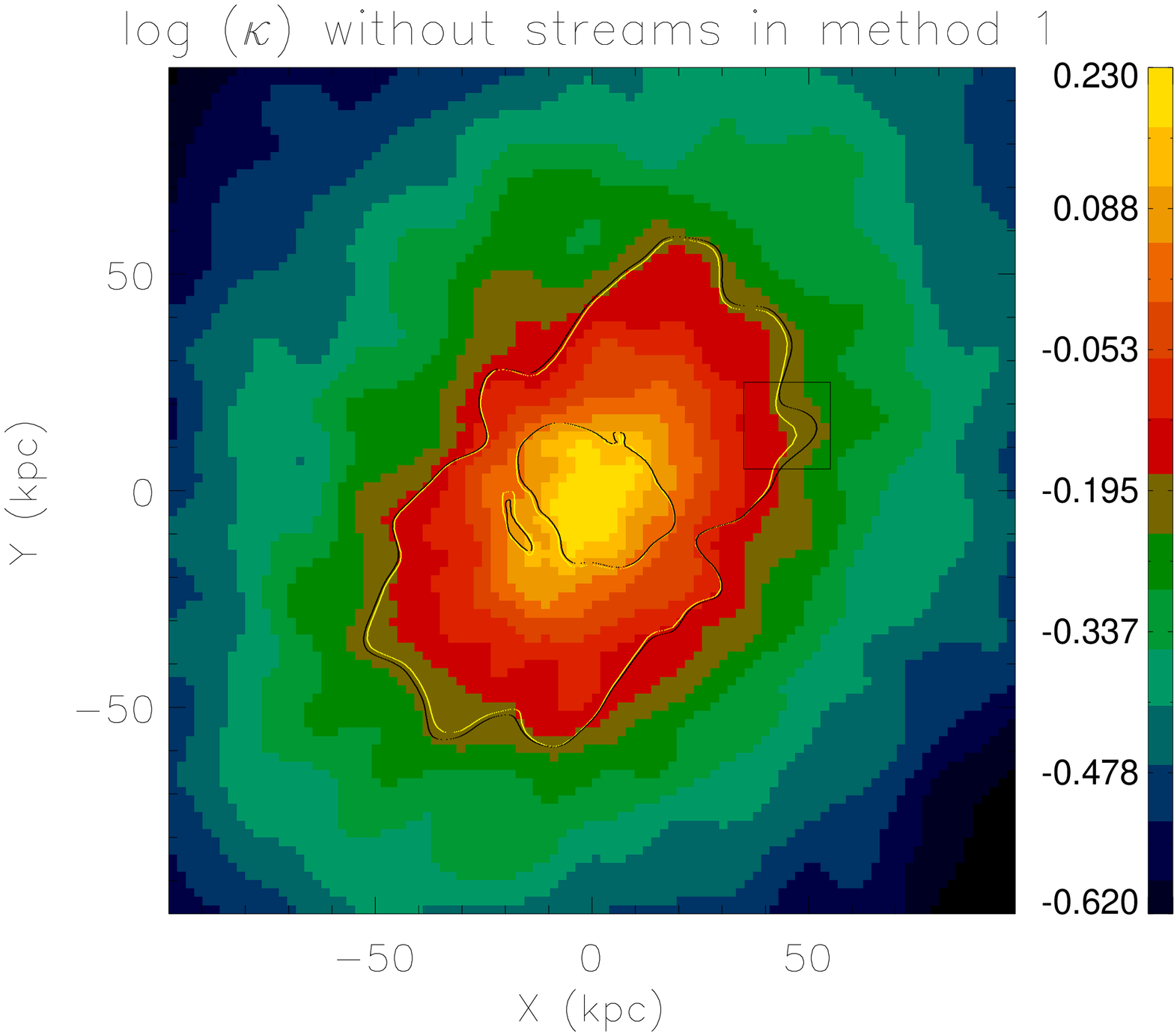}}} 
      \quad
      \subfigure[]{\scalebox{0.35}{\includegraphics{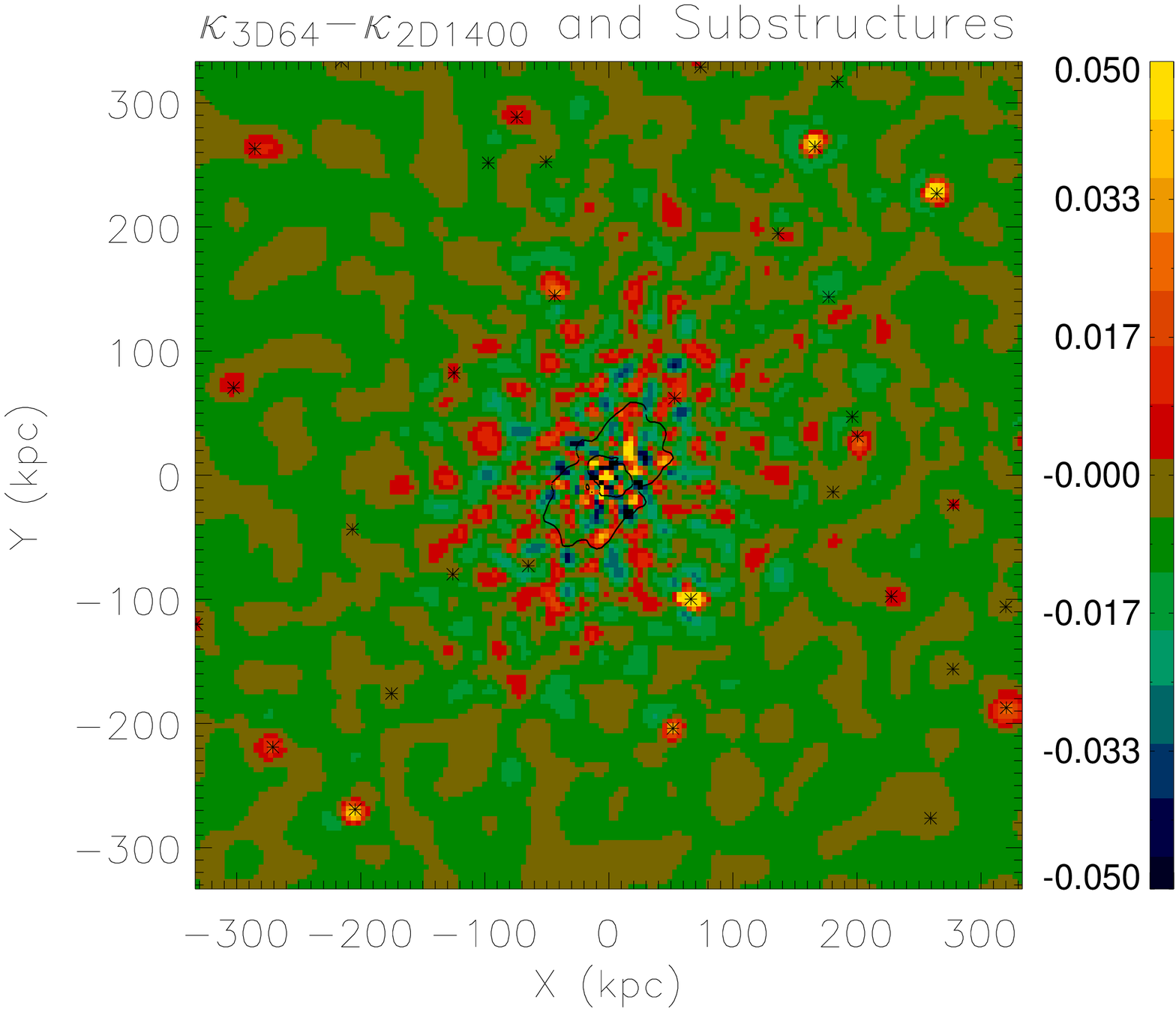}}} 
      \quad
    }
  \end{center}
  \caption{Surface density contours and critical curves (in black and yellow ) for the lensing cluster. a) for
    all the particles. To facilitate the comparison between the different cases, the critical curves obtained with all particles are repeated as the black curves 
in the other three panels. b) without bound substructures found by {\tt
      SUBFIND}. c) without bound substructures and satellite streams
    (see Fig. \ref{fig:streams}) identified in the phase space. d) The difference between the
 dimensionless surface density obtained using  the SI
method and the 2D SPH method. The star symbols indicate the position
of bound substructures. Notice that the SI method usually gives larger
surface densities at the positions of the substructures.
}
  \label{fig:contours}
\end{figure}

\begin{figure}
  \begin{center}
    \mbox{
      \subfigure[]{\scalebox{0.3}{\includegraphics{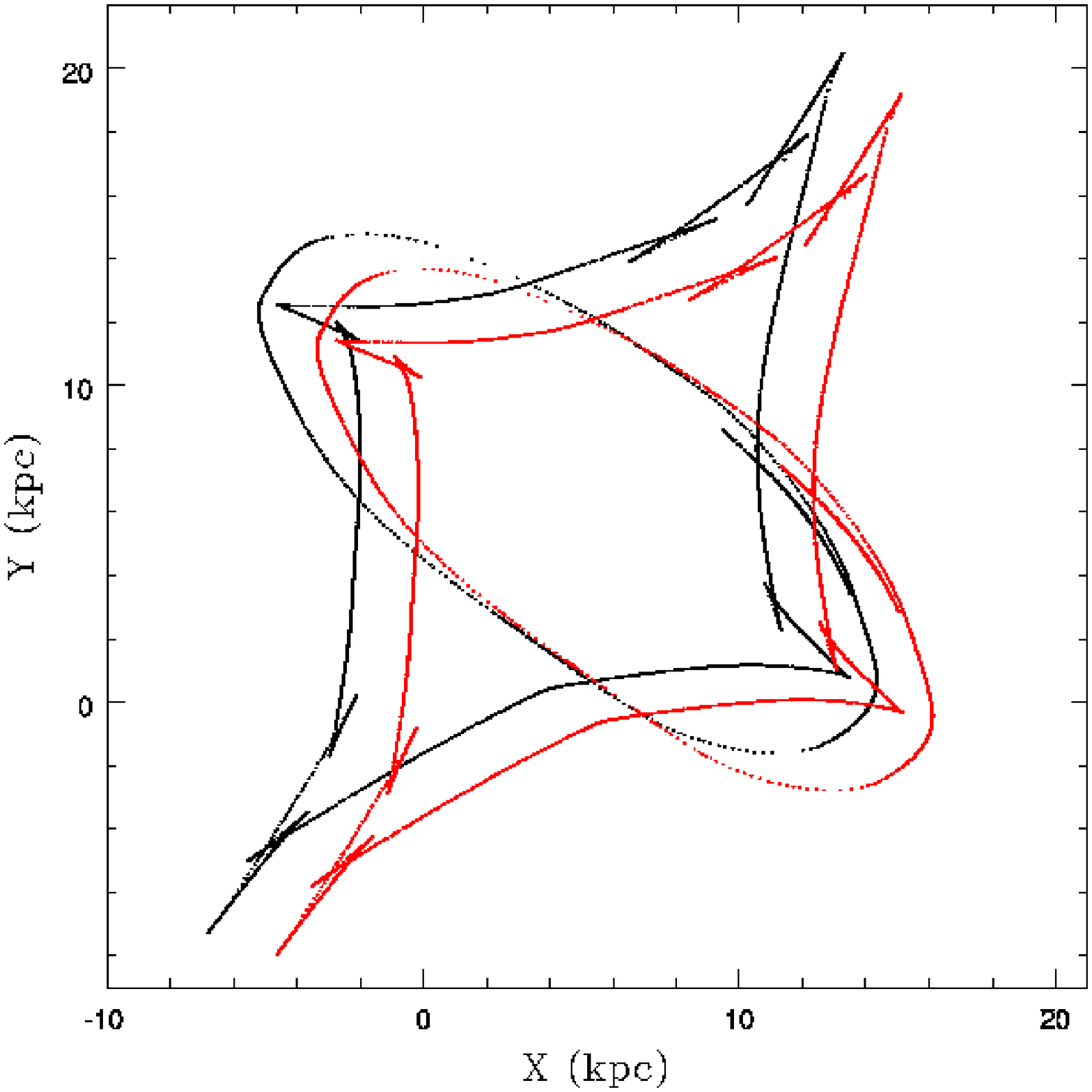}}} 
      \quad
     }
    \mbox{
      \subfigure[]{\scalebox{0.3}{\includegraphics{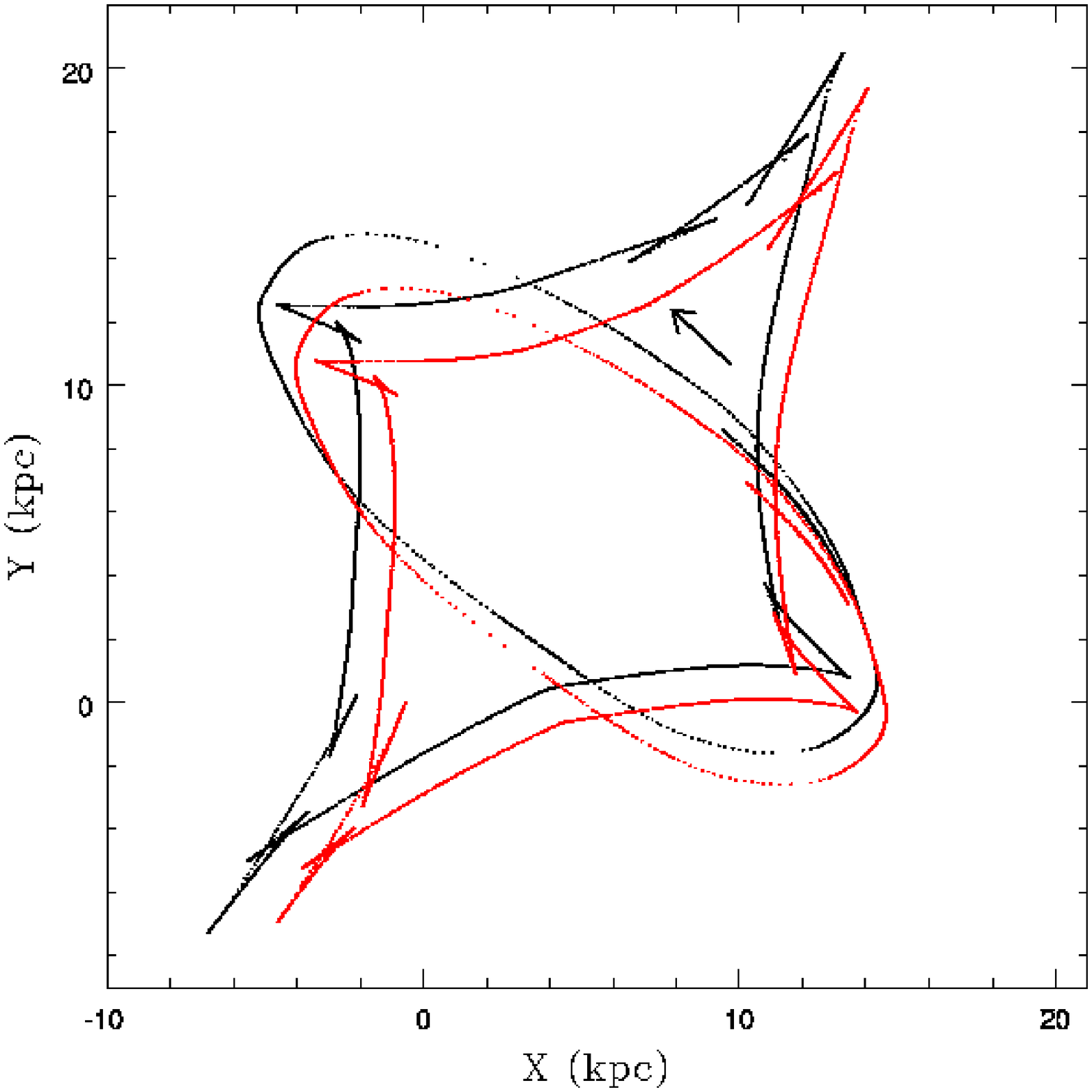}}} 
      \quad
    }
    \mbox{
      \subfigure[]{\scalebox{0.3}{\includegraphics{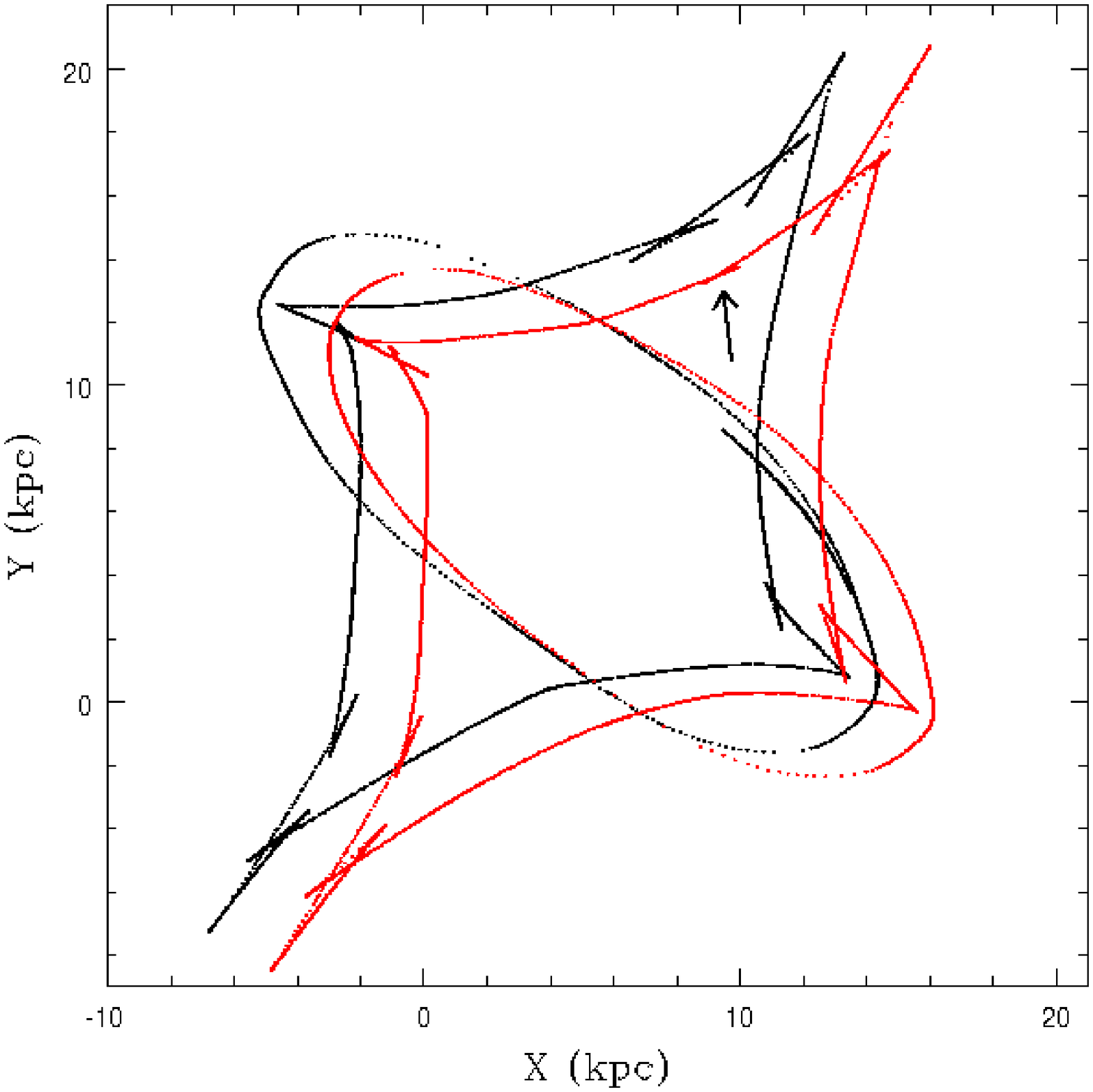}}} 
      \quad
     }
    \mbox{
      \subfigure[]{\scalebox{0.35}{\includegraphics{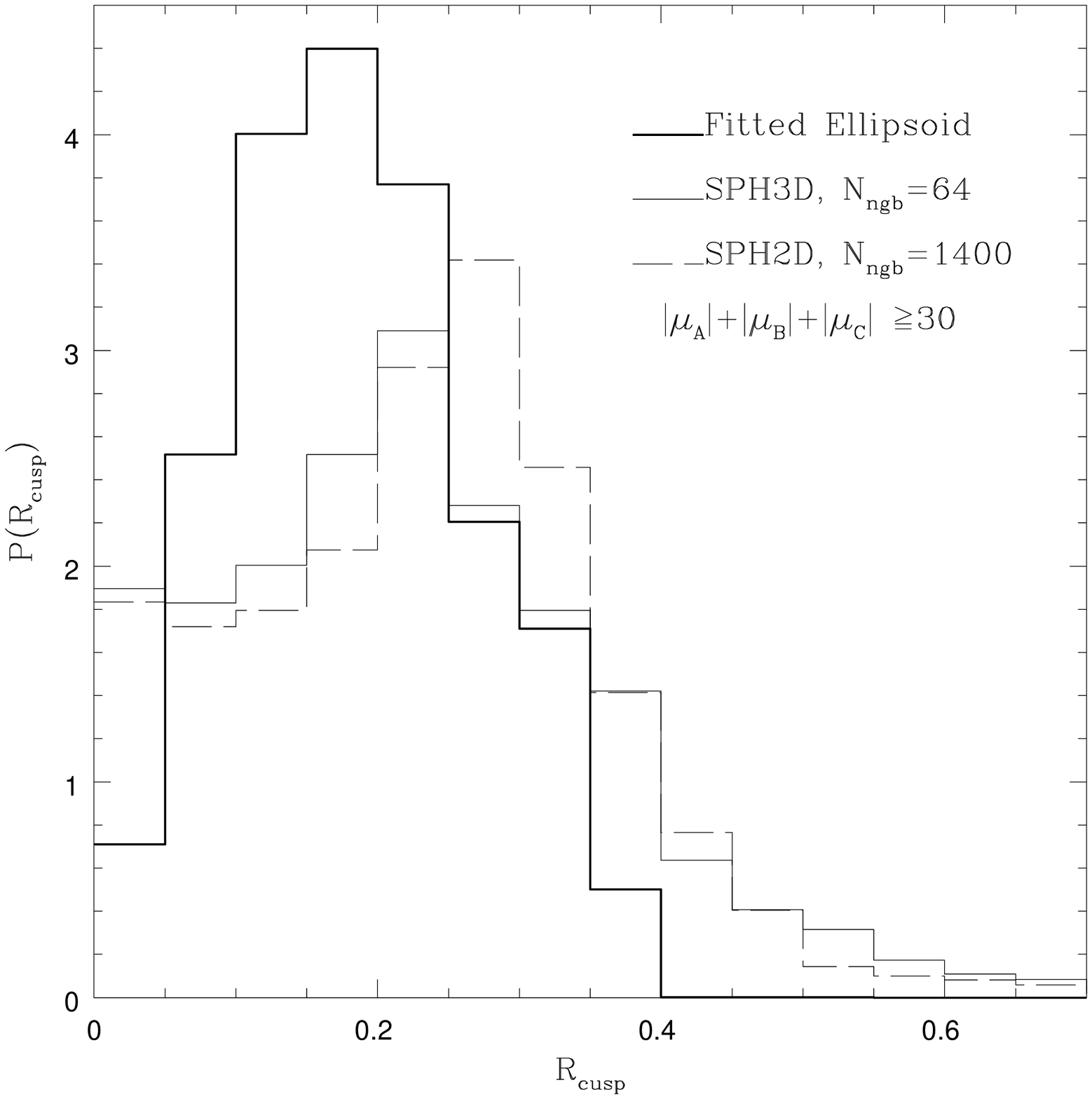}}} 
      \quad
    }
  \end{center}
  \caption{Caustic structures and the probability distribution for
    $R_{\rm cusp}$. For the top left, right and bottom left panels,
the black caustics are obtained using all the particles.
The red lines in the two top panels are the corresponding caustics of
the top right and bottom left panels in
Fig. \ref {fig:contours}, where we exclude the bound substructures and
streams. The top left panel indicates that the substructures in this lens have no 
effect on high-order singularity regions. The top right panel shows the
effect of streams (the most changed region is indicated by the small
arrow). A high-order singularity has disappeared because a
stream in the position of the small box in the bottom left 
panel in Fig \ref {fig:streams} has been removed.  The bottom left panel
shows the caustics from the 2D SPH method. Here, the high-order
singularity also disappears due to the over-smoothing of the boxed
region in the 2D SPH method.
The bottom right panel shows the probability distribution of $R_{\rm
cusp}$.  The 2D SPH smoothing method slightly decreases the probability
for high $R_{\rm cusp}$ (see the text).
}
  \label{fig:caustics}
\end{figure}

We now consider the projected surface mass densities with and without
substructures and streams using the SI smoothing method in
Fig. \ref{fig:contours}. The number of neighbors in the SI algorithm is set to 64.
The top panel in Fig. \ref{fig:contours} shows the surface density
contours and critical curves when all the particles are included,
and the corresponding
caustics are shown in the top panel of Fig. \ref{fig:caustics}. The top
right panel in Fig. \ref{fig:contours} shows the surface density contour
when we exclude the bound substructures identified by {\tt
  SUBFIND}. The corresponding caustics are shown in
Fig. \ref{fig:caustics}. Comparisons show that there is
very little difference between the contours
of the two sets of surface densities and caustics. The exclusion of streams,
however, makes one wiggle in the critical curves less prominent (indicated by a
box in the bottom left panel of Fig. \ref{fig:contours}), and this causes one
of the higher-order singularities to disappear. This case can be explained by the stream
at (45, 15) kpc identified in Fig. \ref{fig:streams}.  Most wiggles in  the critical curves and
many perturbations in the surface density plane still remain unchanged.
It thus appears that streams (in addition to substructures) can explain
one of the high-order singularities for this cluster. However,
it looks that the majorities of high-order singularities in
this case are caused by other reasons. One possibility is
large-scale (low-amplitude) fluctuations in the
main halo. Such fluctuations are a natural result of
incomplete virialization in the merging process on cluster scales.

\begin{figure}
  \begin{center}
    \mbox{
      \subfigure[]{\scalebox{0.4}{\includegraphics{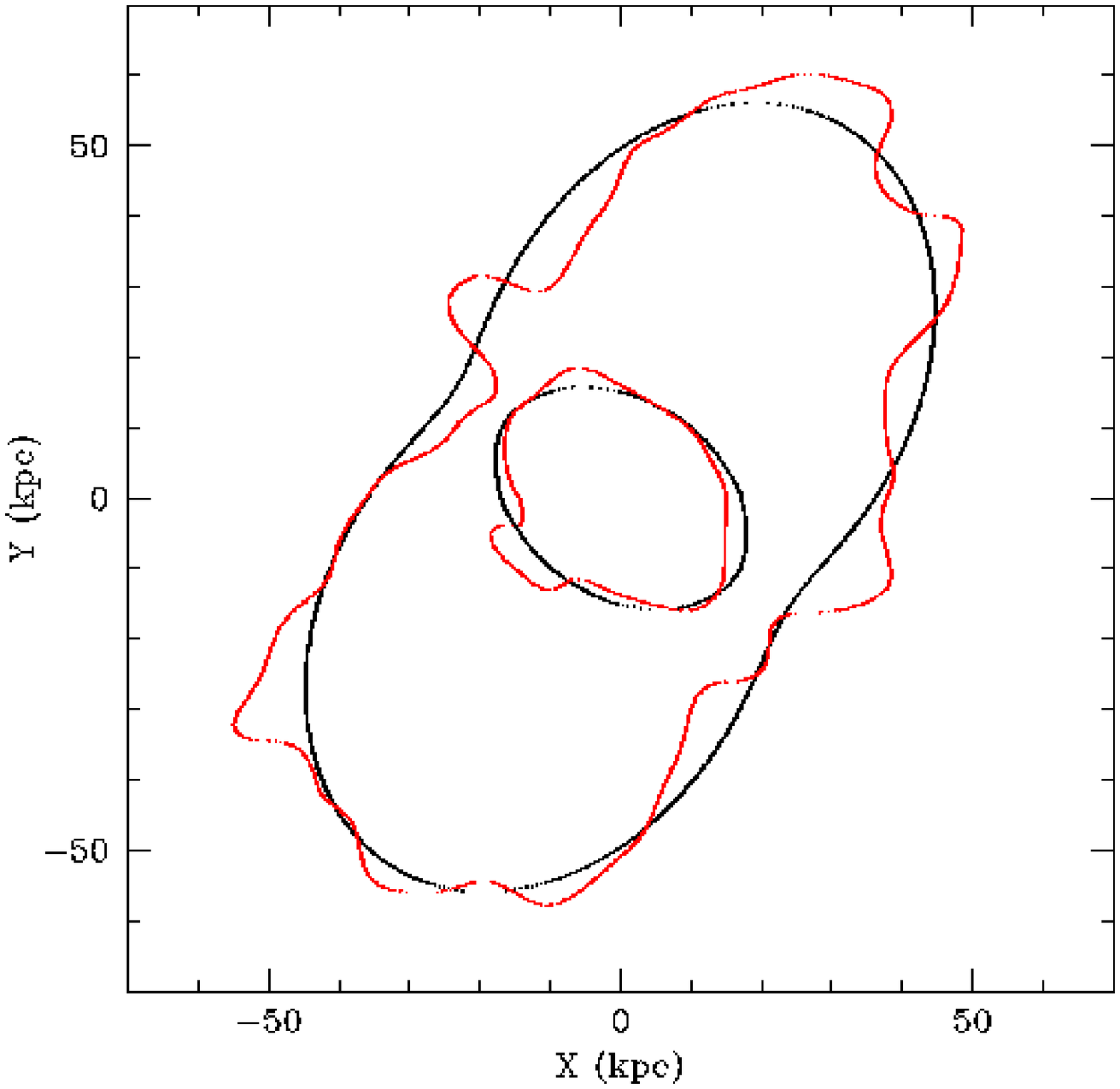}}} 
      \quad
      \subfigure[]{\scalebox{0.4}{\includegraphics{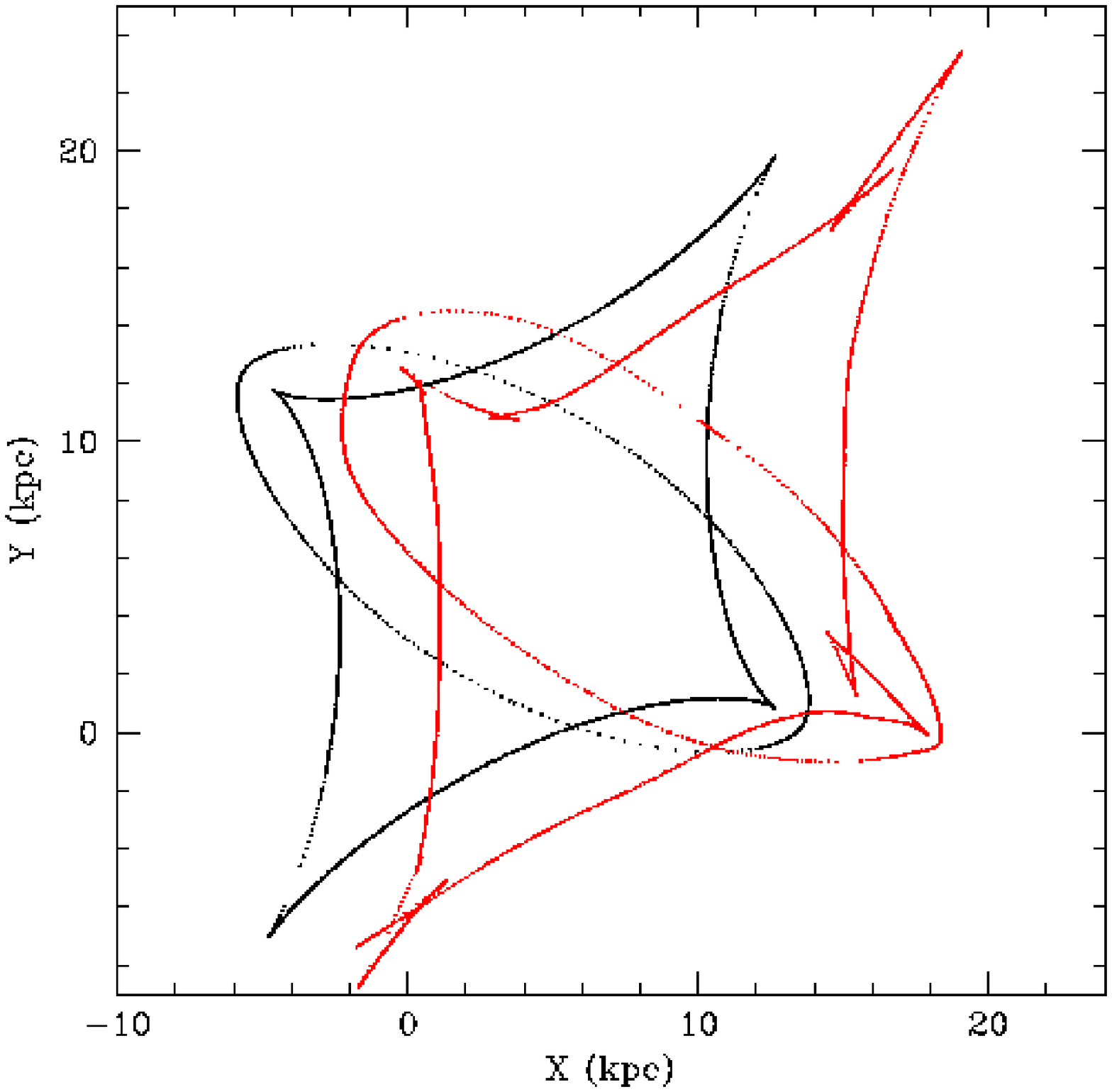}}} 
      \quad
    }
  \end{center}
  \caption{The critical curve (left) and caustics (right) for various
approximations for the cluster obtained from cosmological
simulations. The black curves in the each panel show the critical curves
and cautics for a smooth representation of the cluster and the red curves
 show that for a Monte Carlo realization of the smooth cluster 
with the same number
of particles as in the cosmological cluster.
A high resolution version is available at 
http://202.127.29.4/mppg/english/data/
}
  \label{fig:ellipse}
\end{figure}

However, this may be difficult to prove, as we show below that particle noise
can also cause similar levels of high-order singularities.
We first fit the density contours as ellipses, and find the radial
density profile as a function of the elliptical coordinate, $R_{\rm e}$, where
$R_{\rm e}^2=x^2+y^2/q^2$ and $q$ is the axial ratio. This creates
a smooth representation of our numerical cluster. The critical
curve and caustics for this fitted elliptical model are shown in Fig.
\ref{fig:ellipse}. There are no high-order singularities for this smooth
model. We then study the effect of shot noise by populating the
fitted smooth elliptical with particles in elliptical annuli. For each
annulus, the simulated Monte Carlo cluster has the same number of particles as
the numerical one, but the angular positions of particles are randomized.
Fig. \ref{fig:ellipse} shows the resulting
critical curves and caustics for the Monte Carlo resampled cluster.
A number of high-order singularities appear which are clearly
due to discrete particle noise (about 4.5\% in our case).
We return to the issue of noise at the end of this section.

One key objective of a good smoothing algorithm is the ability to
maintain the information of substructures. In the bottom panel of Fig. \ref{fig:contours},
we show the difference in the obtained surface densities for the 2D and
3D SPH smoothing algorithms. To
keep roughly the same noise level, 1400  neighours are used in the 2D SPH smoothing
based on the condition found by equating eqs. (\ref{eq:SI}) and (\ref{eq:2d}).
Following the discussions in \S\ref{sec:uniform} and \S\ref{sec:SIE},
the noise level in the SI method will in general be lower than that
in the 2D SPH smoothing method as the SI method reduces noise due to chance
alignments in projection.
To show the difference more clearly,
pixels with values outside the limits of the color bar
are set equal to the upper or lower limit. We can clearly see that
most bound structures identified by {\tt SUBFIND} are located in the
positive region, i.e., the SI method picks out these overdense
substructures more reliably, i.e., the 2D SPH algorithm may have over-smoothed.
We believe this also applies to the central region. There are relatively few
bound substructures around the critical curve because most substructures are destroyed by the
tidal force. As we mentioned before, we traced the evolution of the cluster
out to redshift unity and found that in total 88 progenitors
eventually merge to produce the cluster -- many of the
particles in these clumps are not yet virialized, as can be seen in the
satellite streams in Fig. \ref{fig:streams}. So
fluctuations (shown as red) in the inner region are real,
highlighting the over-smoothing by the 2D SPH smoothing method.

\subsubsection{Cusp relations}

Sources close to and inside cusps in general will produce three close
images. The properties of such images have been studied theoretically (Blandford \& Narayan 1986;
Blandford 1990; Schneider \& Weiss 1992; Mao 1992 \& Zakharov 1995). The
magnifications of the three close images follow a simple relation:
\begin{equation}
R_{\rm cusp} = \frac{|{\mu_{\rm A} + \mu_{\rm B} + \mu_{\rm C}}|}
{|\mu_{\rm A}| + |\mu_{\rm B}| + |\mu_{\rm C}|}
\rightarrow 0, \; {\rm for} \; \mu_{\rm tot} \rightarrow \infty\;,
\label{eq:rcusp}
\end{equation}
where A, B and C are three images in the cusp system and $\mu_{\rm tot}$ is the
sum of unsigned magnifications of these three images (the denominator in
the equation above).
Cusp images are more sensitive to the presence of
substructures (Mao \& Schneider 1998; Keeton, Gaudi, \& Petters 2003)
than the separation of images or arc numbers. If the lensing potential
is smooth, then  $R_{\rm cusp}$ must vanish asymptotically as a source
gets close to a cusp.
We use the cusp relation to check the difference in the lensing
properties of the 2D and 3D SPH smoothing algorithms.
For a given  surface density, the lensing potential is obtained using the FFT
method (e.g., Bartelmann et al. 1998; Li et al. 2005). A regular grid is put
on the source plane and the
Newton-Raphson method is used to find all the images and their corresponding
magnifications.

For a source position which produces more than three images, we first
determine the polar angles for all the images measured from the cluster center. The opening angle
between any two images with the same parity are measured. The two images
with the smallest opening angle are selected as the two outer cusp images.
The third image can be easily identified as it has the opposite parity
and lies inside the two outer cusp images. We restrict cusp systems to
the lenses where $\mu_{\rm tot}$
is larger than 30. This is a simple way of selecting systems that have three highly-magnified close images.
Fig. \ref {fig:caustics}d shows the probability
distribution of $R_{\rm cusp}$ for the 2D and 3D SPH smoothing algorithms.
 The 2D
SPH method decreases the probability for high $R_{\rm cusp}$ because it
tends to over-smooth the structures compared with the SI method.
However, the most striking result is that the Monte Carlo resampled cluster
also produces large deviations, just as seen in the numerical
simulated cluster, showing that the noise can mimic the high-order
singularities (see Fig. \ref {fig:ellipse}).
The problem is due to the Poisson noise whose
level is still too high, $\sim 4.5\%$.
Notice that, to decrease the noise level, we cannot simply increase the
number of neighbors in the SI smoothing because this may
over-smooth and remove crucial information on substructures and voids.

\begin{figure}
  \epsscale{0.8} 
  \plotone{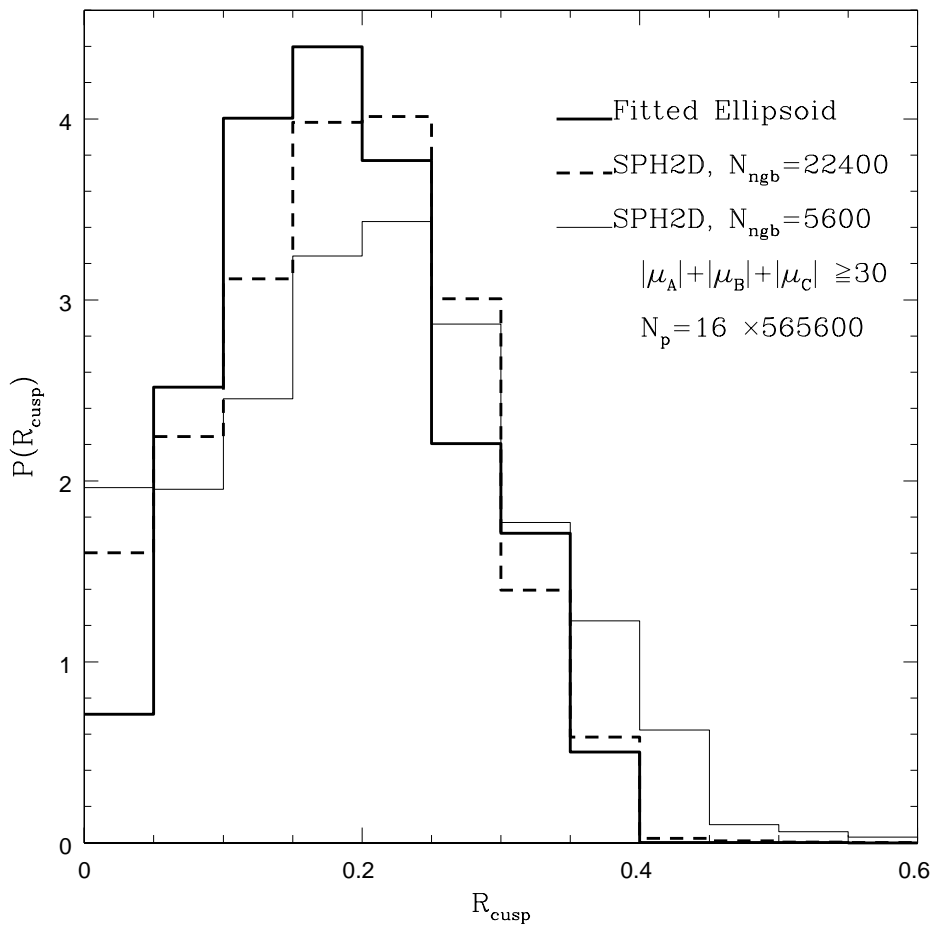}
  \caption{The  effect of noise on the cusp relation. The mass
    resolution is increased by a factor of 16 in the Monte Carlo
    realization. The number of neighbors in SPH2D is set as 4$\times$1400=5600
 and 16$\times$1400=22400. From
eq. (\ref{eq:SI}), the noise levels are approximately 2.2\% and 1.1\% respectively. 
The distribution for the latter reproduce the theoretical cusp relation
for the smooth elliptical (solid line).
}
  \label{fig:Rcusp16}
\end{figure}

Of course, we expect the cusp relation to be better satisfied if we increase the
number of particles in our volume and approach the limit of a smooth potential. To verify this, we generated a
new cluster for which we increase the number of particles by a
factor of 16. The mass of each particle is lowered by the same factor so
as to keep the total mass of the cluster the same.
If we take the $N_{\rm ngb}$ in the 2D SPH smoothing kernel
to be $4 \times 1400$, then the predicted relation still
shows large deviations from the smooth elliptical density distribution
(see Fig. \ref {fig:Rcusp16}). However, when $N_{\rm ngb}$ are
increased by a factor of 16, the probability distribution of $R_{\rm cusp}$
resembles that of a perfect smooth elliptical distribution. From
eq. (\ref{eq:SI}), the noise level drops to about $\approx 1\%$. In this
case, the level of noise appears to be low enough to reproduce the
theoretically expected cusp magnification relation.
According to our comparisons of surface densities with different noise levels,
this value is approximately the highest tolerable level for reliably
reproducing the cusp relation for this cluster. This means that, in order to
suppress the impact of noise or quantify the effect of (sub)structures
on the cusp relation, the mass resolution should be increased by a
factor of 16. In our case, the particle number, $N_p$, should be
increased to $\sim 1\times 10^{7}$ or more. Furthermore, $N_{\rm ngb}$ sets a threshold
for the smallest (sub)structures that we can resolve
reliably. Substructures with $\ga \Nb$ particles will be over-smoothed. In other words, for a given
lower mass limit of (sub)structures above which we
we want to investigate the cusp relation, the particle
number needed in simulations can be very approximately estimated using
eq. (\ref{eq:SI}) for a desired surface density fluctuation.
The best $\Nb$ needs to be chosen on the
 balance of the lowest mass of clumps we want to resolve and the
noise level one can tolerate.

\subsection{Comparisons with the Delaunay tessellation field estimator \label{sec:comparison}}

Bradac et al. (2004) presented a detailed study of the lensing properties of a numerically
simulated elliptical galaxy (Meza et al. 2003). They used
the Delaunay tessellation method in three dimensions to obtain the surface density maps. They also added
an external shear and calculated the resulting critical curves, caustics
and the cusp relation (cf. eq. \ref{eq:rcusp}).
In this subsection, we present comparisons between their results and ours
for this galaxy.

\begin{figure}
{
 \centering
 \leavevmode
 \columnwidth=.45\columnwidth
 \includegraphics[width={\columnwidth}]{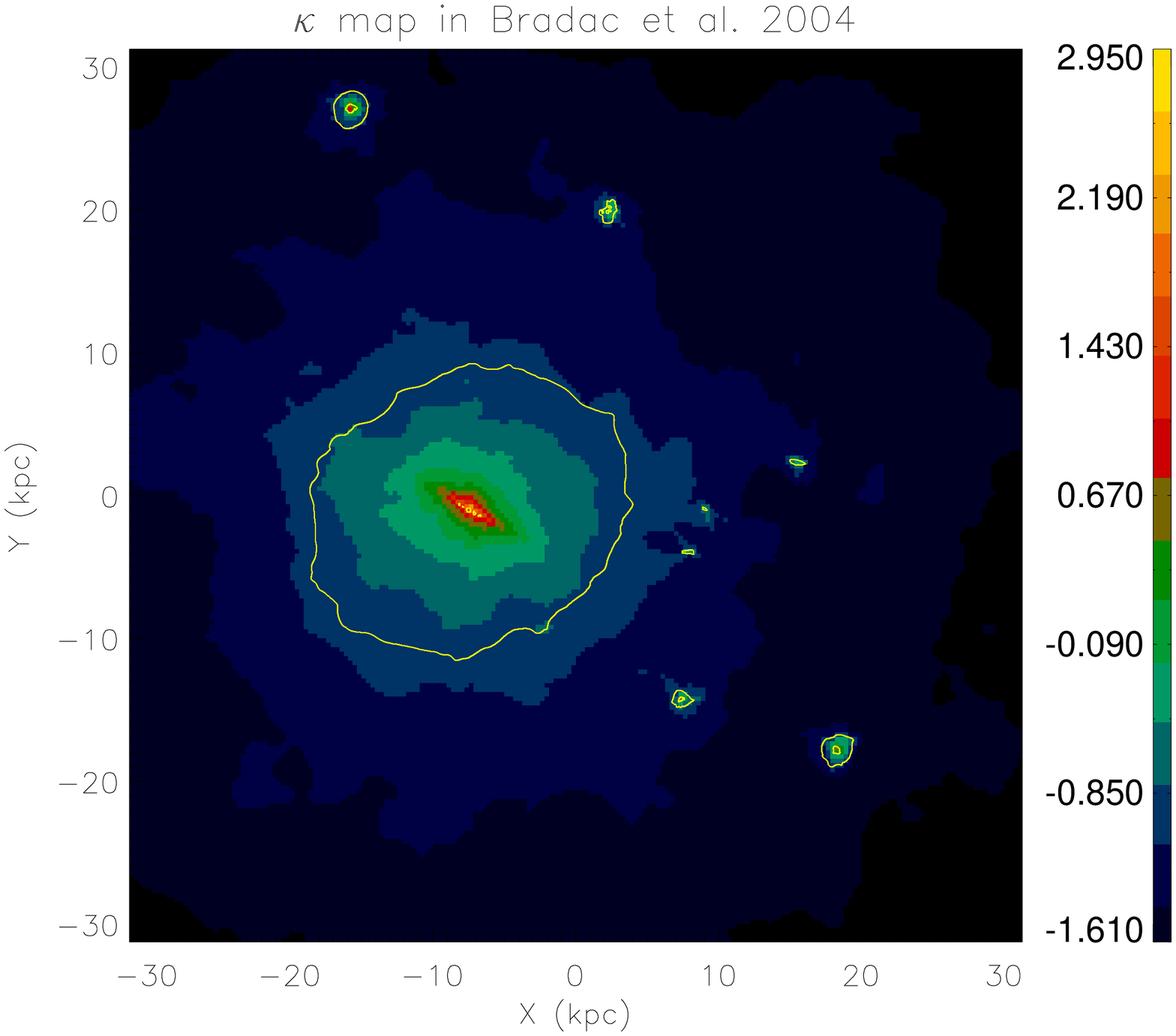}%
 \hfil
 \includegraphics[width={\columnwidth}]{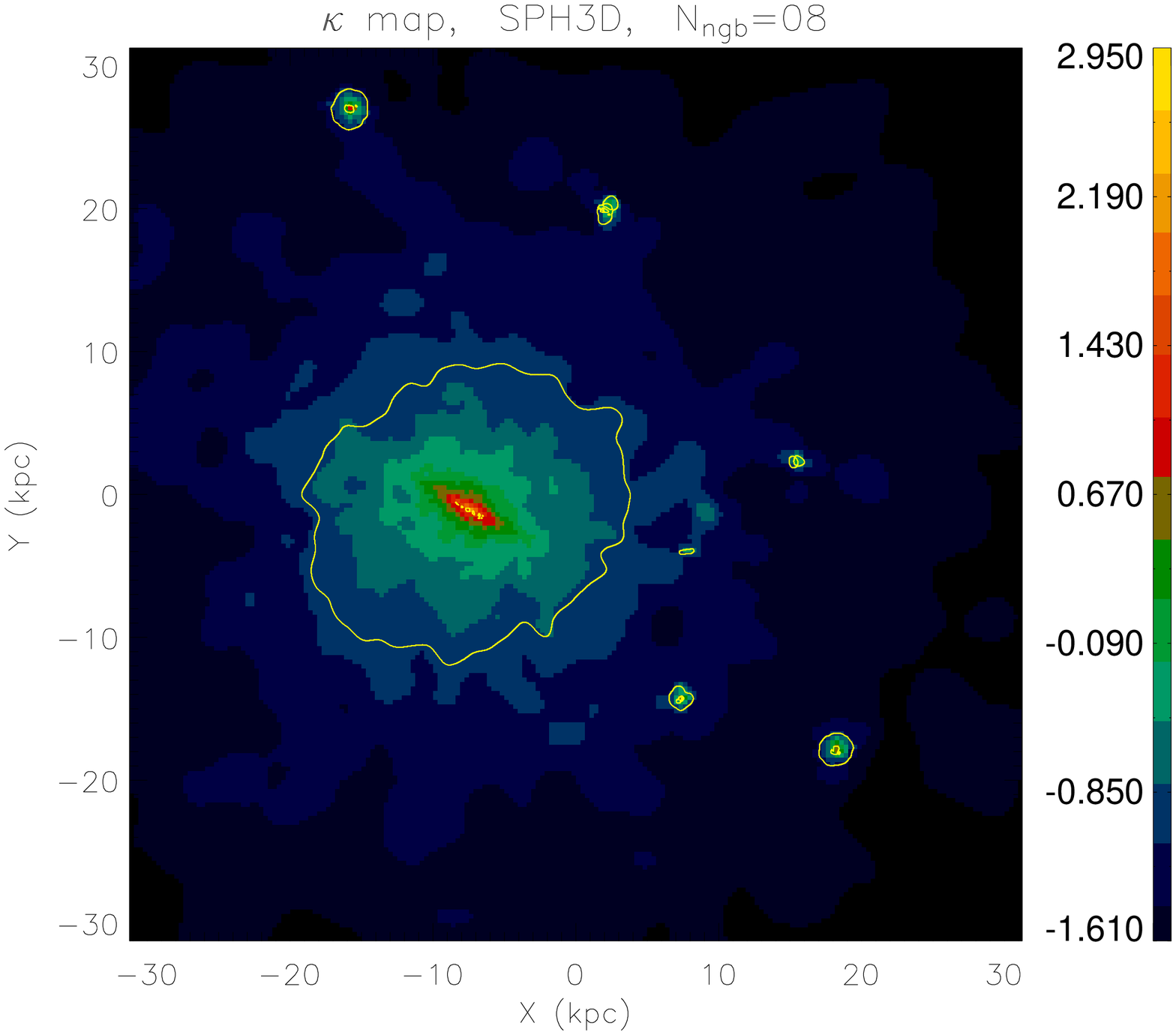}%
\\
 \includegraphics[width={\columnwidth}]{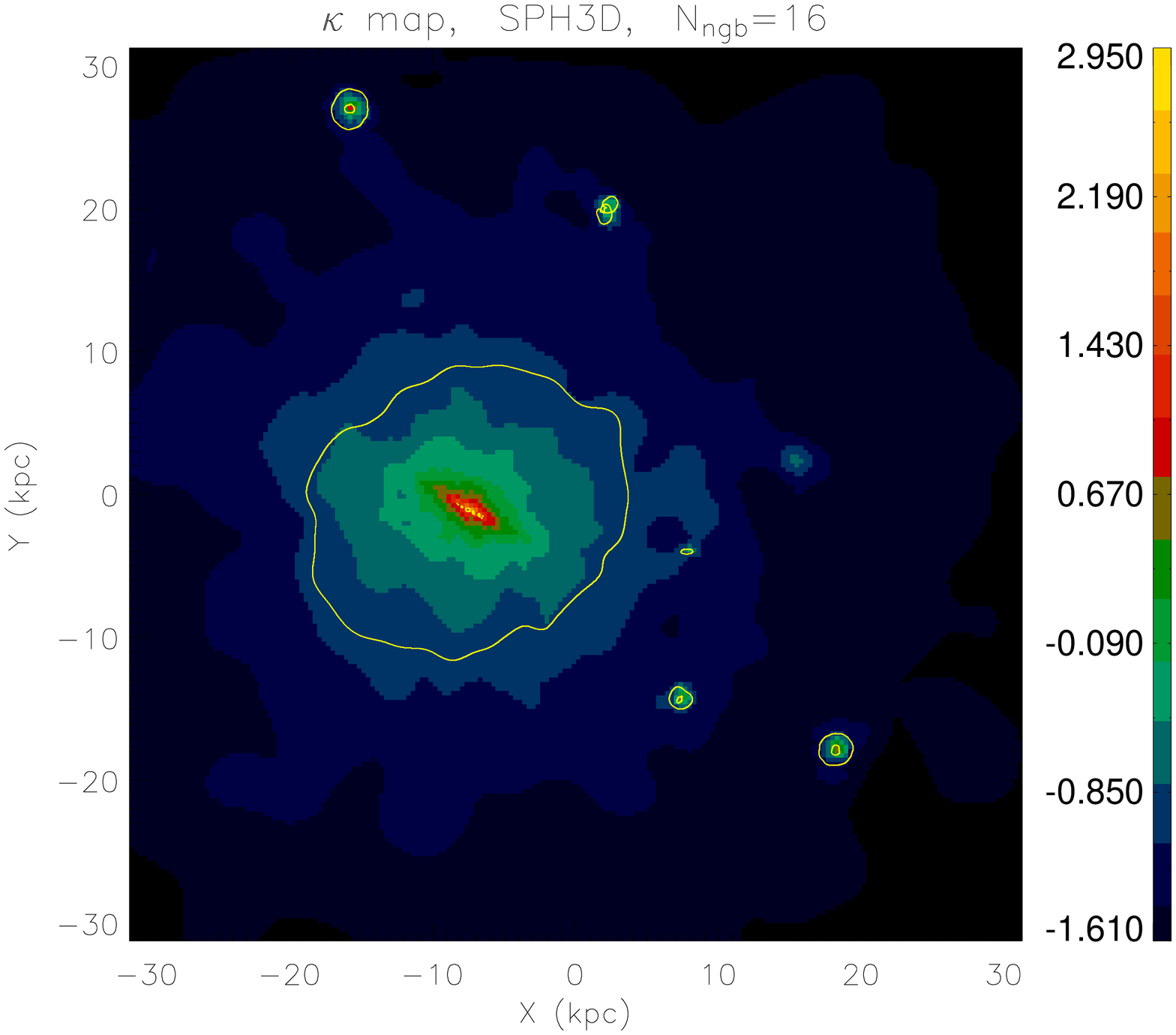}%
 \hfil
 \includegraphics[width={\columnwidth}]{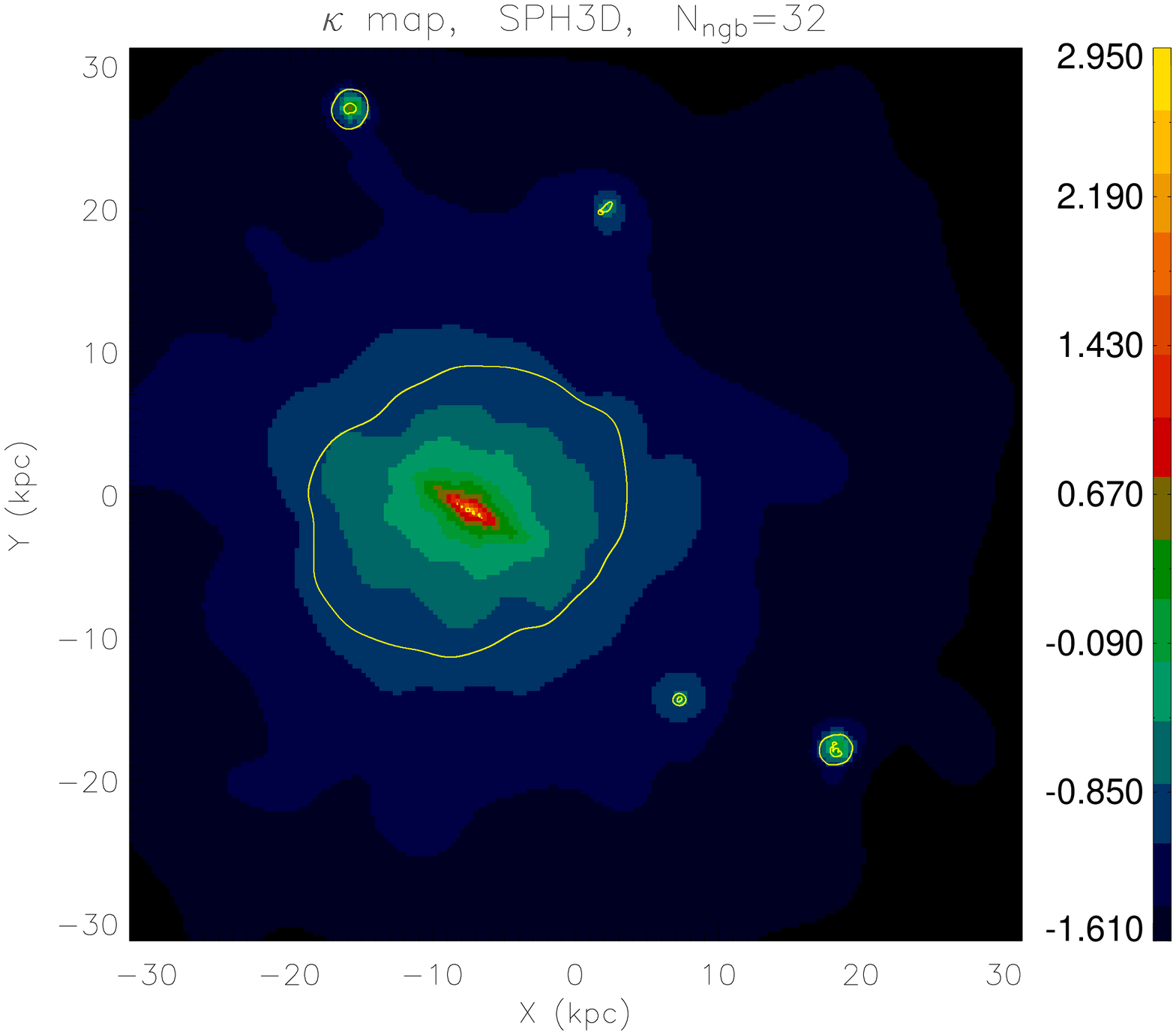}%
}
  \caption{The surface density maps obtained using the
Delaunay tessellation field estimator (Bradac et al. 2004, top left
panel), and those with the SI method with $\Nb=8, 16$ and 32. The solid
(yellow) line in each panel shows the critical curve.
}
\label{fig:bradacKappa}
\end{figure}

\begin{figure}
  \epsscale{0.8} 
  \plotone{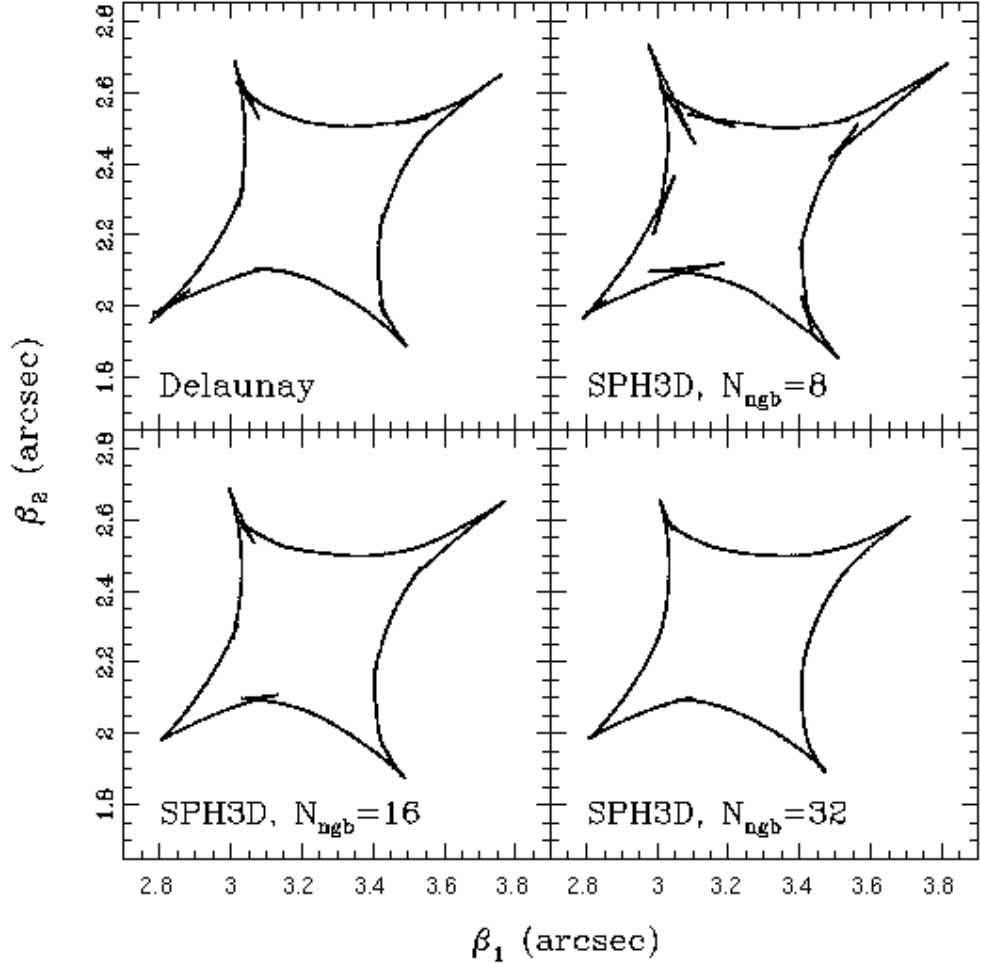}
  \caption{ Caustics corresponding to the surface density maps shown in Fig. \ref{fig:bradacKappa}.
The top left figure is in good agreement with Fig. 2a in 
Bradac et al. (2004) obtained using the Delaunay tessellation field estimator. The
noise in this map is comparable to that in the SI method for $\Nb=16$
in the bottom left panel.}
  \label{fig:bradacCaustics}
\end{figure}

The top left panel of Fig. \ref{fig:bradacKappa} presents the
surface density map and the critical curve for the galaxy
projected along the $z$-axis by Bradac et al. (2004). The other three panels
show the surface density maps obtained using the SI method
with $\Nb$=8, 16 and 32. It is clear that the maps are similar,
particularly the one with $\Nb=16$ seems to match the Bradac et
al. surface density map (the top left panel), including the substructures.
The $\Nb=32$ surface density map may be somewhat over-smoothed, as
the clumps appear circular.
The corresponding caustics for the four surface density maps are shown
in Fig. \ref{fig:bradacCaustics}. For the $\Nb=16$, the caustics match that
obtained through their Delaunay tessellation method quite well.
In short our method yields results that are consistent with those obtained
with the Delaunay tessellation method.

Bradac et al. (2004) also assessed the reality of high-order singularities
and the violation of the cusp relation seen in their simulated galaxy.
To do this, they created an ellipsoidal galaxy with a power-law density profile,
 $\rho(r)\propto r^{-2.9}$. The steep slope is adopted to match that
seen in the simulated cluster around the outer critical curve. Their
simulated galaxy shows no higher-order singularities
and violations of the cusp relation (see their Fig. 8). This is different from
our Monte Carlo simulation (see Fig. \ref{fig:Rcusp16}) where violations of the cusp
relation (and high-order singularities) seem to persist.

\begin{figure}
  \epsscale{0.8} 
  \plotone{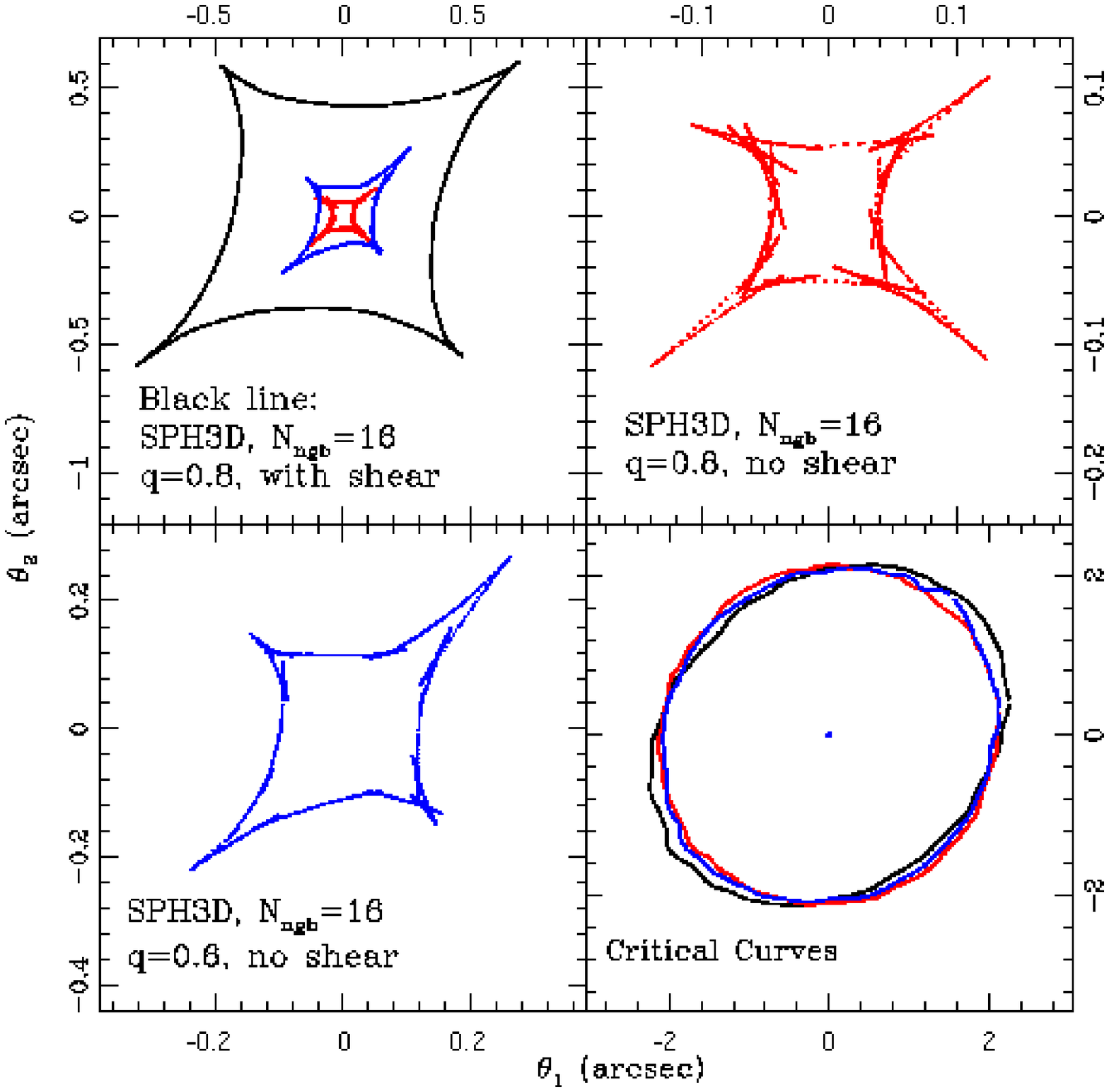}
  \caption{The effect of shear and ellipticity on the caustics and
high-order singularities for different realizations of an oblate
galaxy projected along one of the major axes.
All the galaxies have about 95000 particles within the same box of
Bradac et al. (2004) and the density profile is a power-law, $\rho(r) \propto r^{-2.9}$.
The major axis has an angle of 45$^{\circ}$ with respect to the
horizontal axis. For the top right and bottom left panels the axis ratios
are 0.8 and 0.6 respectively. The black curve in the top
left panel has an axis ratio of 0.8 and an external shear (with the
same value as in Bradac et al. 2004).
Notice how the sizes of caustics and high-order singularities change.
In particular, for the black curve (in the top left panel) there are no high-order singularities
(see text).
}
\label{fig:comp_shear}
\end{figure}

To understand the behavior seen in their simulations, we created several
Monte Carlo realizations of a prolate galaxy for different axis ratios and external shear.
We use the same number of particles ($\sim 95000$) within the
virialized region as they used and the surface density is obtained
along the projection of one of the major axes. For the first realization, we use
an axis ratio 0.8 and the main axis has an angle of $45^{\circ}$ with
respect to the horizontal axis. In this case, the caustics are small and
show many high-order singularities (see the top right panel in
Fig. \ref{fig:comp_shear}). If the axis ratio is changed to 0.6 (the bottom
left panel), the caustics are a factor of two larger, but the high-order
singularities are less important. If we add an external shear (as in Bradac
et al. 2004) to the case with an axis ratio of 0.8, then we obtain
similar caustics (the black-line in the top left panel of Fig. \ref{fig:comp_shear})
as the one shown in their Fig. 8a. Notice that the caustics are much larger and
there are no more high-order singularities. In all three realizations, the
shapes of their critical curves are rather similar (see the bottom panel).
This exercise demonstrates that high-order singularities are not
only a function of the surface density fluctuations but also the
detailed form  of the smooth component of the lens potential, including the ellipticity and shear.

A more quantitative understanding can be achieved using the results
in Evans \& Witt (2001) where they modeled the lensing
potential as an isothermal sphere with arbitrary angular dependences
(see also Witt, Mao \& Keeton 2000). They showed that for galaxies
with higher ellipticities (which usually have larger caustic sizes)
require a higher level of fluctuations to exhibit higher order
singularities (see their Fig. 3 and eq. 32). We believe the difference
between our results and those in Bradac et al. can be understood
due to the difference in the adopted potential forms.
the external shear in their Monte Carlo simulation may act as
 a large effective ellipticity which enlarges the caustics relative to those
seen in our Monte Carlo simulations with ellipticities 0.5 and 0.7.
As a result, the importance of high-order
singularities may be suppressed with similar fluctuations in the surface
densities.

This also explains the difference between Figs. 5 and. 9. The noise levels around the
critical curves are quite similar in these two cases (both are around
5\%).  However, there are no visually striking singularities in the caustics
for the isothermal ellipsoid case while for the simulated cluster,
higher-order singularities are quite prominent. This arises because for the
SIE cluster, the axis ratio is 0.5 while for the numerically simulated
cluster, the axis ratio is larger (0.7). The caustic size for the SIE
cluster is correspondingly larger. The difference in the axis ratio and
caustic sizes means that the same level of fluctuation has different effects on the high-order
singularities.\footnote{The most extreme case is when we have an axis-symmetric
cluster, the tangential caustic size is zero. For this case, any
perturbation will break the symmetry and the caustic area becomes non-zero.}
We verified that if we adopt a more spherical isothermal cluster (increasing $q_3$ close to 1),
the caustics becomes smaller. When the axis ratio and caustics are similar to that of the simulated
cluster, then at the same noise level, high-order singularities also appear.

\label{sec:nbody}
\section{Summary and Discussions}

We have used SPH smoothing algorithms in both 2D and 3D to obtain
the surface densities of galaxy clusters. We used  Monte Carlo simulations of a uniform
density field, an isothermal ellipsoid and an $N$-body numerical
cluster to compare these two SPH smoothing algorithms. We find that
in general the 3D smoothing algorithm is
superior to the smoothing of the 2D projected particle distribution.
In particular, our scatter and integrate
method, for which we first scatter the particle mass in the nearby
grid points in 3D and then integrate to obtain the surface density, appears
to reproduce the underlying surface density well. With $\Nb=32$ or 64,
no prominent artificial higher-order singularities are produced for the
isothermal ellipsoid we simulated.

These high-order singularities are particularly important for the
anomalous flux ratio problem. As shown by Bradac et al. (2004),
these often appear close to the cusps, and produce large
cross-sections which violate the cusp relation. For our high-resolution
numerical cluster, we have critically examined how these high-order
singularities arise. It appears that one of the higher order
singularities is related not to bound substructures but to a satellite
stream identified in the phase space. For the ten most massive clusters
in our $N$-body simulation,  we found that there are always one or more compact unbound
structures in the inner region of clusters. So the effects of
streams may be important for cluster lensing.

Our results show that the importance of high-order singularities will be
a sensitive function of the cluster ellipticity and external shears, as
also demonstrated by Evans \& Witt (2001; see Keeton, Mao \& Witt 2000). This
implies that the maximum noise level allowed to produce reliable predictions will
depend on the detailed form of the lensing potential. Galaxies with
small ellipticities are particularly sensitive to the shot noise, and hence
will require many more particles to make reliable predictions.

If the high-order singularities are due to true large-scale fluctuations
induced, for example, by merging, then their presence is sensitive to the
merging state of clusters. As the merger rate likely increases
at higher redshift and for the most massive clusters, it will be
interesting to examine how these higher-order
singularities change as a function of redshift and cluster mass.
These higher-order singularities provide
regions that can form sextuplet and octuplet images (Keeton, Mao \& Witt
2000; Evans \& Witt 2001). Evans \& Witt (2001) estimated that perhaps
$\sim 1\%$ of galaxy-scale lenses may be sextuplet and octuplet imaged
systems. This fraction may be higher for clusters because they
 have much more
irregular mass distributions. Upcoming large surveys  (e.g., Wittman et
al. 2006) will discover a large number of clusters, the chance of seeing these systems
may be quite realistic.

It is however less clear whether such high-order singularities should be
as prevalent as in galaxies. This is because most lensing galaxies
are ellipticals which may have little evolution since redshift of $\sim 1-2$.
So the substructures in galaxies may be more effectively destroyed.
As even state-of-art simulations cannot reproduce galaxies realistically
in a cosmological setting (e.g. Kawata 2001; Westera et al. 2002;
 Adadi et al. 2003 ), the importance of bound subhaloes
and, particularly, streams in numerical simulations remain to be seen.

\acknowledgments
We thank Dr. Y. Ascasibar for helpful discussions and the referee for an
insightful report.  We are indebted to Dr. M. Bradac for providing us
the data for the numerical galaxy presented in \S\ref{sec:comparison} and many
helpful discussions.
The research in Shanghai is
supported by NKBRSF (G19990754), by NSFC (Nos. 10125314, 10373012, 10303004),
 and by Shanghai Key Projects in Basic research (No. 04jc14079).
SM and LGL acknowledge partial financial support of Chinese Academy
of Sciences and the European Community's Sixth Framework Marie
Curie Research Training Network Programme, Contract No.
MRTN-CT-2004-505183 ``ANGLES".

{}


\begin{references}
\reference{} Abadi, M. G., Navarro, J. F., Steinmetz, M., \& Eke, V. R. 2002, \apj, 591, 499
\reference{} Ascasibar, Y., \& Binney, J. 2005, MNRAS, 356, 872
\reference{} Bartelmann, M. 2003,  gravitational lensing winter school, Aussois, la vieille Europe
(astro-0304162)
\reference{} Bartelmann, M., Huss, A. Colberg, J. M.,
    Jenkins, A., \& Pearce, F. R. 1998, A\&A, 330, 1 (B98)
\reference{} Bartelmann, M., \& Weiss, A. 1994, \aap, 284, 285
\reference{} Bartelmann, M., Steimetz, M. \& Weiss, A. 1995, \aap, 297, 1
\reference{} Bartelmann, M., Meneghetti, M., Perrotta, F., Baccigalupi, C., \& Moscardini, L. 2003, A\&A, 409, 449
\reference{} Blandford, R., \& Narayan, R. 1986, \apj, 310, 568
\reference{} Blandford, R. D. 1990, QJRAS, 31, 305
\reference{} Bradac, M., Schneider, P., Steinmetz, M., Lombardi, M.,
    King, L. J., \& Porcas, R. 2002, \aap, 388, 373
\reference{} Bradac, M., Schneider, P., Lombardi, M., Steinmetz, M.,
    Koopmans, L. V. E., \& Navarro, J. F. 2004, \aap, 423, 797
\reference{} Dalal, N., Holder, G., \& Hennawi, J. F. 2004, \apj, 609, 50
\reference{} Evans, N. W., \& Witt, H. J. 2001, \mnras, 327, 1260
\reference{} Helmi A., White S. D. M., \& Springel V. 2003, MNRAS, 339, 834
\reference{} Hernquist, L., \& Katz, N. 1989, ApJS, 70, 419
\reference{} Hockney, R. W., \& Eastwood, J. W. 1981, Computer Simulation
    Using Particles (McGraw-Hill: New York)
\reference{} Jing, Y. P., \& Suto Y. 2002, ApJ, 574, 538
\reference{} Kawata, D. 2001, \apj, 558, 598
\reference{} Keeton, C. R., Kochanek, C. S. 1998, \apj, 495, 157
\reference{} Keeton, C. R., Mao, S., \& Witt, H. J. 2000, \apj, 537, 697
\reference{} Keeton, C. R., Gaudi B. S., \& Petters A. O., 2003, \apj, 598, 138

\reference{} Kochanek, C. S., 1991, \apj, 373, 354
\reference{} Kochanek, C. S., \& Dalal, N. 2005, \apj, 610, 69
\reference{} Kochanek, C.S., Schneider, P., Wambsganss, J., 2004,
Gravitational Lensing: Strong, Weak \& Micro, Proceedings of the 33rd
Saas-Fee Advanced Course, G. Meylan, P. Jetzer \& P. North,
eds. (Springer-Verlag: Berlin)
\reference{} Kormann, R., Schneider, P., \&  Bartelmann, M. 1994, A\&A, 284, 285
\reference{} Li, G. L., Mao, S., Jing, Y. P., Kang, X., Bartelmann, M., \& Meneghetti, M. 2005, ApJ, 635, 795
\reference{} Lombardi, M., Schneider, P., \aap, 392, 1153
\reference{} Macci\`o, A. V., Moore, B., Stadel, J., Diemand J. 2006, \mnras, 366, 1529
\reference{} Macci\`o, A. V., Miranda, M. 2006, \mnras, 368, 599
\reference{} Mao, S. 1992, \apj, 389, 63
\reference{} Mao, S., \& Schneider, P. 1998, \mnras, 295, 587
\reference{} Mao, S., Jing, Y. P., Ostriker, J. P., \& Weller, J. 2004, \apj, 604, 5
\reference{} Meneghetti, M., Bolzonella, M., Bartelmann, M.,
    Moscardini, L., \& Tormen, G.. 2000, \mnras, 314, 338
\reference{} Meneghetti, M., Bartelmann, M., \& Moscardini, L.
    2003, \mnras, 346, 67
\reference{} Meneghetti, M., Bartelmann, M., Dolag, K., Moscardini, L.,
    Perrotta, F., Baccigalupi, C., \& Tormen, G. 2004, New Astronomy
    Reviews, 49, 111
\reference{} Meza, A., Navarro, J. F., Steinmetz, M., Eke, V. R. 2003, ApJ, 590, 619
\reference{} Monaghan, J. 1992, ARAA, 30, 543
\reference{} Navarro, J. F., Frenk, C. S., \& White, S. D. M. 1997,
 \apj, 490, 493
\reference{} Niedereiter, H. 1978, Bull. Amer. Math. Soc, 84, 957
\reference{} Puchwein, E., Bartelmann, M., Dolag, K., \& Meneghetti,
M. 2005, \aap, 442, 405
\reference{} Schaap, W.E., \& van de Weygaert, R. 2000, A\&A, 363, L29
\reference{} Schneider, P., \& Weiss, A. 1992,A\&A, 260, 1
\reference{} Springel, V., White, S. D. M., Tormen, G., \& Kauffmann, G.
2001, \mnras, 328, 726
\reference{} Torri, E., Meneghetti, M., Bartelmann, M., Moscardini, L.,
     Rasia, E., \& Tormen, G. 2004, \mnras, 349, 476
\reference{} Wambsganss, J., Bode, P., \& Ostriker, J. P.,
\reference{} van de Weygaert, R. 1994, A\&A, 283, 361
2004, ApJ, 606, L93 (W04)
\reference{} Westera, P., Samland, M., Buser, R., \& Gerhard, O. E. A\&A, 389, 761
\reference{} Wittman, D.,  Dell'Antonio, I. P., Hughes, J. P.,
     Margoniner, V. E., Tyson, J. A., Cohen, J. G., \&  Norman, D. 2006, \apj, 643, 128
\reference{} Zakharov, A. F. 1995, A\&A, 293, 1
\end{references}
\end{document}